\documentclass[]{emulateapj}

\def\ie{i.e.}
\def\eg{e.g.}
\def\degree{\hbox{$^\circ$}}

\def\kms{\hbox{km$\;$s$^{-1}$}}
\def\ms{\hbox{m$\;$s$^{-1}$}}

\shorttitle{Magneto-acoustic waves in roAp stars}

\shortauthors{Khomenko \& Kochukhov}

\begin{document}

\title{Simulations of magneto-acoustic pulsations in atmospheres of rapidly oscillating Ap stars}

\author{E. Khomenko\altaffilmark{1,2}, and O. Kochukhov\altaffilmark{3}}
\email{khomenko@iac.es}

\altaffiltext{1}{Instituto de Astrof\'{\i}sica de Canarias, 38205,
C/ V\'{\i}a L{\'a}ctea, s/n, Tenerife, Spain}
\altaffiltext{2}{Main Astronomical Observatory, NAS, 03680, Kyiv,
Ukraine} \altaffiltext{3}{Department of Physics and Astronomy,
Uppsala University, Box 515, SE-751 20, Sweden}

\begin{abstract}
Rapidly oscillating Ap stars exhibit an astrophysically
interesting combination of strong, dipolar-like magnetic fields
and high-overtone  {\it p-}mode pulsations similar to the Sun.
Recent time-resolved spectroscopy of these stars unravelled a
complex picture of propagating magneto-acoustic pulsation waves,
with amplitude and phase strongly changing as a function of
atmospheric height. To interpret these observations and gain a new
insight into the atmospheric dynamics of roAp stars we have
carried out 2-D time-dependent, non-linear magneto-hydrodynamical
simulations of waves for a realistic atmospheric stratification of
a cool Ap star. We explore a grid of simulations in a wide
parameter space, treating oscillations of the velocity, magnetic
field and thermodynamic quantities in a self-consistent manner.
Our simulations foster a new understanding of the influence of the
atmosphere and the magnetic field on the propagation and
reflection properties of magneto-acoustic waves, formation of node
surfaces, and relative variation of different quantities. Our
simulations reproduce all main features of the observed
pulsational behavior of roAp stars. We show, for the first time,
that the overall depth dependence of the pulsations in roAp
atmospheres is strongly influenced by the density inversion at the
photospheric base.

\end{abstract}



\section{Introduction}

\subsection{General properties of roAp stars}
\label{sec:genprop}

Rapidly oscillating Ap (roAp) stars are main sequence, late-A
chemically peculiar stars showing high-overtone {\it p-}mode
pulsations with periods between 6 and 22 minutes
\citep{kurtz:2000,kochukhov:2008c}. Effective temperatures of
these stars span a relatively narrow range from 6400 to 8100~K.
Their atmospheres exhibit large abundance anomalies, typical of
the group of SrCrEu chemically peculiar stars. The light and
iron-peak elements are normal or underabundant relative to the
chemical composition of the solar photosphere, whereas
concentration of the rare-earth elements (REEs) is enhanced by
factors $10^3$--10$^5$ \citep{ryabchikova:2004a}.

The presence of very strong, organized magnetic fields is another
remarkable property of roAp stars. The mean field strength
inferred from the polarization measurements and Zeeman splitting
of spectral lines usually lies in the 1--5~kG range
\citep{mathys:1997b,Hubrig+etal2004,kochukhov:2006} but can reach
25--30~kG in exceptional cases
\citep{kurtz:2006b,freyhammer:2008}, making Ap stars the most
extreme non-degenerate magnetic stars known. These fields have
simple, dipolar-like topology, often significantly inclined with
respect to the stellar rotation axis
\citep{landstreet:2000,bagnulo:2002}. Unlike rapidly evolving,
dynamo-generated magnetic fields of the Sun and late-type stars,
the fields of Ap stars are most likely stable remnants of the
fields acquired in the pre-main sequence evolutionary phase
\citep{moss:2004,braithwaite:2006}. Ap stars also rotate
significantly slower than their normal counterparts
\citep{abt:1995}, probably due to an enhanced braking provided by
a magnetized wind and a more efficient coupling with the
circumstellar disk in the pre-main sequence stage
\citep{stepien:2000}.

The unusual atmospheric chemistry of Ap stars is attributed to the
effects of atomic diffusion \citep{michaud:1970}. This slow
process of chemical separation under the influence of competing
forces of the radiative pressure and gravity is facilitated by the
slow rotation and stabilizing effect of the strong magnetic field.
Different diffusion velocities in the regions permeated by the
horizontal and vertical magnetic field lines lead to substantial
variation of chemical abundances over the surfaces of magnetic A
stars \citep{michaud:1981,alecian:2004}. These chemical starspots
are responsible for the spectacular rotational modulation of line
profiles in the spectra of roAp stars, synchronized with the
variation of the overall stellar brightness, photometric colors,
magnetic field strength and the mean line of sight magnetic field
component \citep{ryabchikova:1997a,kochukhov:2004e}.

Chemical stratification in the line-forming atmospheric regions is
another important consequence of the atomic diffusion in Ap stars.
Detailed empirical line profile studies
\citep{bagnulo:2001,ryabchikova:2002} and self-consistent
theoretical diffusion calculations \citep{leblanc:2009} show that
the light and iron-peak elements are concentrated in the lower
atmosphere layers, typically below continuum optical depth at
$\lambda=5000$ \AA\ $\log\tau_{5}=-1:0$. On the other hand, REEs,
such as Pr and Nd which show many strong lines in the roAp
spectra, are pushed by the radiative pressure to the optically
thin layers above $\log\tau_{5}=-3:-4$
\citep{mashonkina:2005,mashonkina:2009}.

According to the current consensus, roAp pulsations are driven by
the $\kappa$ mechanism operating in the \ion{H}{1} partial
ionization zone with the additional influence of chemical
gradients and magnetic quenching of convection
\citep{balmforth:2001,theado:2009}. The efficiency of this
excitation mechanism sensitively depends on the relationship
between the magnetic field, convection, and atomic diffusion. The
physics of these processes is not entirely understood.
Consequently, despite giving robust and informative pulsation
frequency predictions
\citep[e.g.,][]{cunha:2003,gruberbauer:2008}, the current roAp
excitation theories are unable to account for the H-R diagram
distribution of 40 known roAp stars and cannot explain their
difference from apparently constant but otherwise very similar
non-pulsating Ap stars.

Shortly after discovery of the {\it p-}mode pulsations in Ap stars
a close connection between oscillations and magnetic field became
obvious. Pulsational amplitude and phase were observed to follow a
rotational modulation curve correlated with the field strength
variation. The times of the magnetic and pulsation maxima
generally coincide, while pulsation phase shows a characteristic
$\pi$ radian jump when the stellar magnetic equator is closest to
the line of sight \citep[e.g.,][]{kurtz:1994a}. This behavior has
inspired the {\it oblique pulsator model} \citep{kurtz:1982},
which attributes roAp pulsation to a non-radial, axisymmetric, low
angular degree modes aligned with the axis of dipolar magnetic
field. Theoretical calculations \citep{Saio2005} and indirect
imaging of the horizontal pulsation velocity field structure
\citep{kochukhov:2004f} have vindicated the oblique pulsator model
but also revealed that the modes are substantially distorted by
the magnetic field and cannot be described by a single spherical
harmonic function.


\subsection{Spectroscopic pulsation signatures of roAp stars}

The advent of efficient, high-resolution spectrographs at large
optical telescopes allowed extending observational analysis of
roAp pulsations from traditional high-speed, broad-band photometry
\citep[see a thorough review by][]{kurtz:2000} to radial velocity
(RV) measurements and studies of the individual line profile
variability \citep[see reviews
by][]{Kochukhov2007,kochukhov:2008c,Kurtz2008}. These
investigations revealed a number of unusual atmospheric pulsation
phenomena, not present in any other type of pulsating star and not
anticipated theoretically.

The most prominent and unique characteristic of the rapid
spectroscopic variability of roAp stars is a large diversity of
the pulsation amplitudes and phases of different spectral lines
\citep{kanaan:1998,kochukhov:2001b,mkrtichian:2003,elkin:2005,
kochukhov:2006a,ryabchikova:2007b}. The narrow cores of the
hydrogen Balmer lines and absorption features of the rare-earth
ions usually exhibit RV amplitudes from a few hundred \ms\ to
several \kms. On the other hand, lines of Ca, Si  and iron-peak
elements vary with amplitudes below $\sim$\,50~\ms. This amplitude
discrepancy is frequently accompanied by a substantial difference
in the pulsation phases. In extreme cases the phase lags between
RV curves of different lines reach $\pi$--2$\pi$ radians
\citep{mkrtichian:2003,ryabchikova:2007b,sachkov:2008}.

Equally puzzling is the pulsational bisector variation in many
roAp stars. Time-series measurements performed at different
intensity levels in the profiles of strong REE lines and H$\alpha$
core yield systematically different results
\citep{sachkov:2004,kurtz:2005,ryabchikova:2007b,Ryabchikova+etal2007},
typically showing a large increase of the pulsation amplitude
towards the line wings and occasional change of the pulsation
phase across line profiles.

These spectroscopic pulsation signatures of roAp stars are
difficult to reconcile with the oscillation picture expected for a
normal, chemically-homogeneous stellar atmosphere.
\citet{ryabchikova:2002} suggested that peculiar pulsation
properties of roAp stars are related to the presence of strong
atmospheric chemical composition gradients, for which there is
unequivocal evidence from observations and theory alike. Chemical
stratification is particularly extreme for the rare-earth ions,
which levitate so high in the atmosphere that their formation
heights exceed those of the hydrogen line cores. Pulsation waves
propagate outwards in this chemically segregated atmosphere,
gradually increasing in amplitude and consecutively perturbing the
layers probed by the Fe-peak element lines, H$\alpha$ and REE
lines. Thus, time-resolved spectroscopic observations of roAp
stars provide a vertical cross-section of {\it p-}modes and allow
detailed study of the propagation and transformation of
magneto-acoustic waves over a large range of geometrical heights.

Several time-resolved spectroscopic studies suggested that the
phase-amplitude behavior of roAp pulsations show characteristics
of standing waves (nearly constant phase) in deeper layers and
running waves (changing phase) higher in the upper atmosphere
\citep[e.g.,][]{kurtz:2003}. A few roAp stars exhibit node-like
surfaces where the pulsation amplitude drops below the detection
threshold and the pulsation phase undergoes a $\pi$ radian jump
\citep{ryabchikova:2007b,sachkov:2008}.

High-quality spectroscopic time-series obtained for the strongest
REE lines \citep{kochukhov:2001b} demonstrated that roAp
pulsational line profile variation is also highly unusual. Weak
lines of the singly ionized Nd and Pr, which form in the lower
part of the REE-enriched cloud, exhibit modulation expected for a
low angular degree, non-radial pulsation. However, the strongest
REE lines originating in the uppermost layers show asymmetric
blue-to-red running waves, which are incompatible with the line
profile variability pattern produced by any non-radial, horizontal
pulsational velocity fluctuation. \citet{kochukhov:2007}
successfully modeled this phenomenon by a combination of the RV
and line width variation. \citet{shibahashi:2008} suggested an
alternative explanation involving very high amplitude shock waves,
unresolved by observations.

The question of the magnetic field variation with pulsation phase
has been a topic of intense recent debate. \citet{hubrig:2004a}
predicted magnetic variability of the order of $\delta B/B\approx
0.1$ but could not reliably detect pulsational field changes in
any of the six roAp stars they surveyed with  low-resolution
spectropolarimetry. \citet{leone:2003} claimed a detection of
$\delta B/B=0.2$ pulsational field modulation in the REE lines of
the roAp star $\gamma$~Equ. However, these results were not
confirmed by \citet{kochukhov:2004c} who established an upper
limit of $\delta B/B=0.02$ for the same star based on a more
extensive observational material. Subsequent searches of the
pulsational variation of magnetic field by
\citet{kochukhov:2004,savanov:2006,Kochukhov+Wade2007} also
yielded null results.

\subsection{Overview of MHD wave calculations for the Sun and roAp
stars}

To understand the complex physics of waves in magneto-atmospheres,
numerical modeling of wave interaction with magnetic field has
become a preferred approach in recent years, with the development
of large computational facilities. It has an advantage over the
analytical models, allowing to include magnetic field
configurations of nearly arbitrary complexity and to study
compound effects of several different physical agents, such as
radiative losses or non-linear effects. For the Sun, where
localized magnetic structures (sunspots, flux tubes) can be nearly
spatially resolved in observations and studied in detail, the
numerical modeling of waves in the photosphere and chromosphere of
these structures has been performed for several decades, starting
from \eg\ \citet{Shibata1983} and exponentially increasing in time
during the last few years \citep[][without touching the literature
on waves in the corona where the physical conditions are
significantly different from those in the lower
atmosphere]{Cally+Bogdan1997, Rosenthal+etal2002, Bogdan+etal2003,
Hasan+Ulmschneider2004, DePontieu+etal2004, Hasan+etal2005,
Khomenko+Collados2006, DePontieu+etal2007, Steiner+etal2007,
Khomenko+Collados2007, Hanasoge2008, Cameron+etal2008,
Khomenko+etal2008, Hasan+Ballegooijen2008, Moradi+etal2009,
Khomenko+etal2009, Parchevsky+Kosovichev2008}. Important physical
conclusions have been reached by these models. It has been
demonstrated that the behavior of waves depends crucially on
whether they propagate in the magnetically dominated region or in
the gas pressure dominating region. The separation between these
two regions is defined by the value of the plasma parameter
$\beta=P_g/P_m$, the ratio between the gas pressure and the
magnetic pressure. In the particular case of waves, a more
adequate parameter is the ratio between the squared characteristic
wave speeds, the sound speed and the Alfv\'en speed,
$c_S^2/v_A^2$. These two parameters are related as
$(c_S^2/v_A^2)/\beta = \gamma/2 =5/6$ (for an ideal monoatomic
gas). In the magnetized stellar atmosphere the ratio $c_S^2/v_A^2$
decreases exponentially with height as the density decreases. In
the solar case a typical situation is to have $c_S$ of the same
order of $v_A$ at the spectral line formation heights in the
photosphere or low chromosphere, depending on the field strength
of the magnetic structure.

In a two-dimensional situation, when only fast and slow MHD modes
exist, the slow mode is mainly magnetic in the region where
$c_S>v_A$ and is mainly acoustic in the case $c_S< v_A$. The fast
mode nature is the opposite. A detailed description of the fast
and slow modes behavior, refraction and mode conversion in the
solar atmosphere can be found in \eg\ \citet{Bogdan+etal2003,
Khomenko+Collados2006}. The slow mode is mostly field-aligned in
the sense that its group velocity is parallel to the local
magnetic field direction. The fast mode can propagate across the
field. To establish the wave type in observations it is important
to verify the height of the $c_S = v_A$ level with respect to the
line formation level. Recently a lot of attention has been drawn
to the phenomenon of mode transformation
\citep{Zhugzhda+Dzhalilov1982, Zhugzhda+Dzhalilov1984b, Cally2006,
Cally+Goossens2007} taking place around the $c_S=v_A$ level.
Applied to the Sun, mode transformation is believed to be
responsible for the wave power reduction observed in sunspots
\citep{Cally+Bogdan1997} and, possibly, for the wave power
redistribution over active regions \citep{Hanasoge2008,
Khomenko2009}. Depending on the inclination of the field with
respect to the wave propagation direction, the transformation can
be complete or not. But in any case, new modes appear after the
transformation and the wave behavior around the $c_S=v_A$ level
can be extremely complex since different modes, propagating up and
down, can interfere. In a three-dimensional situation, the wave
behavior is even more mixed, and our overall understanding is far
from being complete \citep[however, see][]{Cally+Goossens2007}.

Regarding the magnetized atmospheres of the roAp stars, it has
been generally accepted that the magnetic field makes important
effects on oscillation properties \citep[see the recent reviews
by][]{Cunha2007, Kurtz2008}. A number of theoretical studies have
been carried out to advance understanding of the influence of
magnetic field on oscillation eigenfrequencies, eigenfunctions,
and on the structure of the multiplets and frequency spacings
\citep{Dziembowski+Goode1996, Bigot+etal2000, Cunha+Gough2000,
Saio+Gautschy2004, Saio2005, Cunha2006}. These theoretical models
were able to provide a qualitative explanation of the behavior of
the frequency spacings due to the action of the magnetic field
\citep{Cunha+Gough2000, Saio+Gautschy2004}. Continuing the work of
\citet{Cunha+Gough2000}, \citet{Cunha2006} used a variational
approach to calculate the magnetic perturbations to
eigenfrequencies. She assumed that magnetic perturbations are
introduced only in a thin layer around $c_S=v_A$ (which was called
``magnetic boundary layer'') changing the wave vector and
introducing additional phase shifts. Note that a similar picture
was used to explain the behavior of the helioseismic waves in
solar spots, calling the magnetic effects ``surface effects''
\citep{Braun1997, Lindsey+Braun2005a, Lindsey+Braun2005b,
Moradi+Cally2008, Moradi+etal2009, Khomenko+etal2009}.
\citet{Saio+Gautschy2004} and \citet{Saio2005} addressed the same
problem with a different approach, by using expansions of the
eigenfunctions into spherical harmonics \citep[see
also][]{Dziembowski+Goode1996}. In both approaches, it was found
that, depending on its strength, the magnetic field causes cyclic
effects on the pulsation frequencies. Due to the mode conversion
at the magnetic boundary layer a part of the wave energy escapes
to the high-frequency running acoustic waves (slow low-$\beta$
magneto-acoustic modes) or to the slow high-$\beta$
magneto-acoustic mode waves in the interior, producing mode energy
losses that can become important at some particular frequencies.

As was mentioned above, the amplitudes and phases of the observed
waves depend on the atmospheric height in a complex way. This
behavior was largely not understood until the recent studies by
\citet{Cunha2006} and \citet{Sousa+Cunha2008}. Unlike the Sun,
oscillations observed in roAp stars come from a magnetically
dominated region in the most part of the stellar atmosphere. Only
in the deepest observed layers the gas pressure and Lorentz forces
can be of the same order of magnitude for the lowest field
strengths encountered in roAp stars \citep{Cunha2007}. Due to the
direct influence of the Lorentz force, running acoustic waves in
the atmosphere are supposed to become field-aligned and their
acoustic cut-off frequency changes as a function of the magnetic
field inclination as $\omega_c\cos\theta$, where $\omega_c$ is a
cut-off frequency in the absence of magnetic field
\citep{Dziembowski+Goode1996, Cunha+Gough2000}.
\citet{Sousa+Cunha2008} considered an analytical quasi
one-dimensional model of the radial high-frequency pulsations
(above the cut-off frequency) in a polytropic interior model of
roAp stars matched to an isothermal atmosphere. They found that in
the interior running magnetic waves with progressively decreasing
wavelength exist (slow high-$\beta$ magneto-acoustic modes)
together with nearly standing acoustic waves with much larger
wavelength (fast high-$\beta$ magneto-acoustic modes). In the
atmosphere the situation is the opposite: running acoustic waves
with the displacement parallel to the magnetic field coexist with
the nearly standing magnetic waves with displacement perpendicular
to the magnetic field (slow and fast low-$\beta$ magneto-acoustic
modes, respectively). A superposition of the line of sight
projections of both wave components non-trivially depends on the
magnetic field parameters and can possibly produce a complex
spectroscopic pulsation behavior, in qualitative agreement with
observations \citep{Sousa+Cunha2009}.
For oscillations with frequencies below the cut-off frequency, the
losses to the running acoustic waves in the atmosphere are
restricted to latitudes where the inclination of the dipolar
magnetic field is sufficient to lower down the cut-off according
to $\omega_c\cos\theta$. For high-frequency oscillations, only a
part of the energy is removed due to the running acoustic waves in
the atmosphere and they can become trapped, unlike the
non-magnetic case, where high-frequency waves are always
propagating. The losses to the high-frequency running acoustic
waves in the atmosphere were found to be the largest near the
poles of the magnetic dipole where the field is nearly vertical.
The losses to the running magnetic waves in the interior are the
largest for fields inclined around 30 degrees. These results are
similar to those from local helioseismology mode conversion
studies in inclined magnetic fields by \citet{Cally2006} and
\citet{Schunker+Cally2006}. Thus, except for regions near the
poles, the high-frequency wave energy is nearly trapped.
Extrapolating the results of \citet{Sousa+Cunha2008}, waves with
frequencies below the cut-off frequency are expected to be trapped
at and near the poles but able to convert into running acoustic
slow-mode waves with increasing inclination, so that the mode
energy losses would be maximum in an annulus close to the poles
(but not exactly at the poles). In all the cases, trapped waves
concentrate near the magnetic equator. Note, that earlier works
tried to explain the high-frequency wave reflections by an ad hoc
chromospheric temperature rise \citep{gautschy:1998}, which
contradicted observations since no evidence of chromospheres was
found for the cool Ap stars.

The observational evidence for the lack of magnetic field
variation over pulsation cycle seems to be confirmed theoretically
by \citet{Saio2005} who found that the magnetic field pulsations
do not exceed the limit of  $\delta B/B=10^{-5}$. However, these
theoretical calculations relied on an approximate analytical
description of the atmospheric structure and did not extend to the
low optical depths where strongly pulsating lines are formed.

\begin{figure*}
\center
\includegraphics[width=14cm]{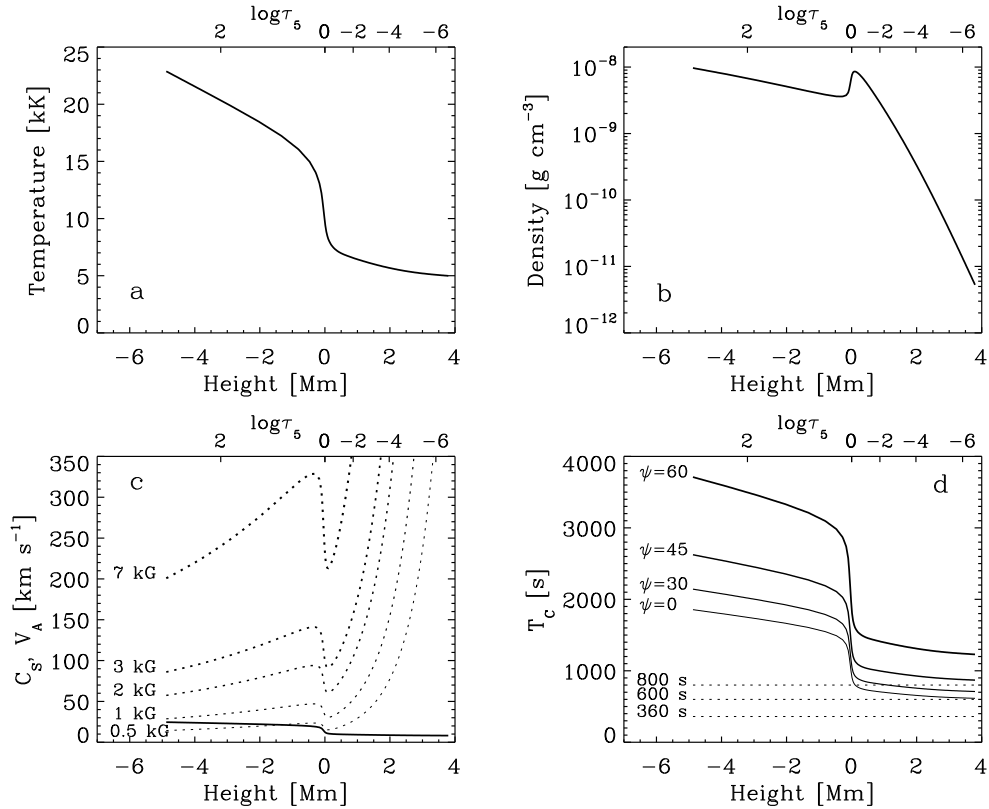}
\caption{Equilibrium model atmosphere. (a) Temperature. (b)
Density. (c) Solid line: acoustic speed; dotted lines: Alfv\'en
speed for the values of the magnetic field marked with numbers.
(d) Solid lines: cut-off period $T_c=2\pi/(\omega_c\cos{\psi})$,
where $\psi$ is the magnetic field inclination and
$\omega_c=\gamma g/2c_S$. Dotted lines: periods of the driver. In
all panels zero height corresponds to log$\tau_5=0$. The optical
depth scale is given by the upper axis. } \label{fig:model}
\end{figure*}

\subsection{Goals of our study}

Despite this recent progress, more theoretical work is needed to
reach the complete understanding of the roAp star's pulsation
properties and, in particular, to explain the unusual
spectroscopic variability of these stars. In this paper we use
two-dimensional numerical simulations to study the properties of
magneto-acoustic waves in the atmospheres of roAp stars. We take
advantage of the numerical code developed by
\citet{Khomenko+Collados2006} to model the wave propagation in
magneto-atmospheres. This code was previously applied to a variety
of problems of solar physics, such as the wave energy transport to
the chromosphere in small-scale magnetic flux tubes, or
helioseismic wave propagation though the magnetic structure of
sunspots \citep{Khomenko+etal2008, Khomenko+etal2009}. Our
strategy is similar to the work of \citet{Sousa+Cunha2008}.
However, since the equations are solved fully numerically, we were
able to relax several approximations made by these authors. We
assume the waves to be excited by a radial pulsation mode below
the visible surface in a stratified non-isothermal atmospheric
model permeated by a constant inclined magnetic field. The code
solves non-linear equations, allowing for development of shocks in
our simulations and treating the propagation of magneto-acoustic
waves in a realistic cool Ap-star atmosphere. This approach allows
us to relate interesting pulsation properties with the formation
heights of important diagnostic lines, thereby bridging the gap
between the complex observational picture of roAp pulsation and
theoretical models, previously limited to simplistic analytical
atmospheric structures.

Our goal is to create a more generalized picture and to study in
detail the effects of the magnetic field strength and inclination,
and the temperature stratification on the observed pulsation
amplitudes and phases. We analyze the behavior of individual
magneto-acoustic modes at different latitude locations and at
different frequencies, both below and above the atmospheric
cut-off frequency.  The signatures of these waves in the
disc-integrated signal are also studied. Our analysis is not
limited just to the velocity oscillations, but we also study
self-consistently the temporal behavior of thermodynamic
parameters and the of magnetic field vector.

Our simulations are suitable for detailed spectral synthesis and a
direct comparison to observations. However, we defer this work to
subsequent papers. Below we concentrate on the analysis of the
physical properties of magneto-acoustic waves under different
conditions and try to address several key questions: why such a
diverse pulsation wave behavior is observed in the atmospheres of
roAp stars, depending on the  spectral line (\ie\ the height in
the atmosphere); which physical mechanisms determine the presence
of the node surfaces, rapid growth of the amplitude and change of
the pulsation phase with height, mixture of an upward and a
downward wave propagation; what kind of temperature, density and
magnetic field variations accompanies RV oscillations, etc.

\section{Local MHD simulations}

The global magnetic field of roAp stars can be approximated to a
first order as a magnetic dipole
\citep{bagnulo:1995,kochukhov:2004e}. The characteristic
wavelength of oscillations can be estimated as $\lambda \sim
c_ST$, where $T$ is the pulsation period. For a typical model
atmosphere of roAp star $c_S$ is approximately a few tens of \kms,
depending on the atmospheric height, and $T$ is of the order of
ten minutes, making the wavelength to be of the order of $10^1$
Mm. This number is negligible compared to the stellar radius and
to the characteristic scale of the variations of the magnetic
field. Based on these considerations we take a local approach in
the simulations, assuming that the magnetic field is locally
homogeneous and inclined.

Given that the observed pulsations are of low degree $\ell$, we
assume that at the base of our simulation domain (below the
visible surface) the oscillations are excited by a plane-parallel
perturbation with mostly vertical displacement.
For the dipolar field and modes with zero azimuthal order (m=0)
the pulsations are axisymmetric.
The simulated waves propagate in a plane defined by the magnetic
field vector and the gravity vector. For the dipolar field the
pulsations are axisymmetric and their properties depend on the
magnetic latitude, but not on the azimuthal angle. Apart from
this, the properties of oscillations in the horizontal direction
are not constrained by any assumption.

\citet{Dziembowski+Goode1996, Cunha+Gough2000, Cunha2006,
Sousa+Cunha2008} also solved a local problem in a plane-parallel
atmosphere with a locally uniform magnetic field, allowing
oscillations to have the displacement vector in the
two-dimensional plane. The difference between our approach and the
approach of these authors is that our analysis is not limited to
low-amplitude oscillations and allows arbitrary variations of the
velocity in horizontal direction in contrast to quasi-radial
pulsations considered in the aforementioned papers. We focus
entirely on the propagation of waves in the stellar atmosphere and
a small portion of sub-surface layers and do not consider
excitation of the pulsations and mode transformation in the
stellar interior.

\subsection{Numerical method}


We use the numerical MHD simulation code that was developed during
the last years at the Instituto de Astrof{\'{\i}}sica de Canarias
(IAC) to study the wave propagation in different magnetic
structures. The description of the code can be found in
\citet{Khomenko+Collados2006, Khomenko+Collados2007}. The code
solves the basic non-linear equations of the ideal MHD, written in
conservative form:
\begin{equation}
 \frac{\partial \rho }{\partial \it{t}} +  \vec{\nabla} \cdot
(\rho \vec{V})=  0 \,, \label{eq:den}
\end{equation}
\begin{equation}
\frac{\partial (\rho \vec{V}) }{\partial \it{t}} + \vec{\nabla}
\cdot [ \rho \vec{V}\vec{V} + (P + \frac{\vec{B}^2}{8\pi}) {\bf I}
- \frac{\vec{B}\vec{B}}{4\pi}]=\rho\vec{g} \,, \label{eq:mom}
\end{equation}
\begin{equation}
\frac{\partial E}{\partial \it{t}} + \vec{\nabla}\cdot[(E + P +
\frac{\vec{B}^2}{8\pi})\vec{V} - \vec{B}(\frac{\vec{B} \cdot
\vec{V}}{4\pi})] = \rho\vec{V}\cdot\vec{g} + \rho Q  \,,
\label{eq:ene}
\end{equation}
\begin{equation}
 \label{eq:ind}
\frac{\partial \vec{B}}{\partial \it{t}} =
\vec{\nabla}\times(\vec{V} \times \vec{B})  \,,
 \end{equation}
where ${\bf I}$ is the identity tensor and $E$ is the total
energy:
\begin{equation}
E=\frac{1}{2}\rho V^2 + \frac{P}{\gamma-1} + \frac{B^2}{8\pi} \,.
 \end{equation}
All other symbols have their usual meaning. In these equations the
gravity $\vec{g}$ is taken constant with a value corresponding to
that at the stellar surface.
In the present version of the code, the radiative energy losses,
$Q$, can be included according to the Newton cooling law. However,
we postpone the study of non-adiabatic effects to the future.

%

The code solves the non-linear MHD equations for perturbations,
that are obtained by subtracting the equations of initial
magnetostatic equilibrium from  Eqs.~\ref{eq:den}-\ref{eq:ind}.
Note that this procedure is not equivalent to a usual
linearization of the equations since we maintain all the
non-linear terms in the equations for perturbations. In the
simulations described above we use the equation of state of an
ideal gas.

The boundary condition issue is very important for wave
simulations since artificial reflections at the boundaries can
distort the physical results and limit the duration of the
simulations. Our code makes use of a Perfectly Matching Layer
(PML) boundary condition \citep{Berenger1994} that efficiently
absorbs the incoming waves at the boundaries and prevent their
return into the simulation domain, using relatively few grid
points. In the simulations described below, the horizontal domain
boundaries are periodic and the top boundary has a PML layer. At
the bottom boundary of the simulation domain we drive sine waves
with a given period by means of the exact analytical solution for
the radial acoustic-gravity waves in an isothermal stratified
atmosphere \citep{Mihalas+Mihalas1984}:
\begin{equation}
V_{z} =V_0 \exp\left(\frac{z}{2H}\right)\sin(\omega t - k_z z)
\label{eq:v}
\end{equation}
\begin{equation}
\frac{\delta\rho}{\rho}=\frac{V_0 k_z}{\omega}|R|
\exp\left(\frac{z}{2H}\right)\sin(\omega t - k_z z + \phi_R)
\label{eq:r}
\end{equation}
\begin{equation}
\frac{\delta P}{P}=\frac{V_0 k_z \gamma}{\omega}|P|
\exp\left(\frac{z}{2H}\right)\sin(\omega t - k_z z + \phi_P)
\label{eq:p}
\end{equation}
where $H=P/(dP/dz)$ is the pressure scale height and the
amplitudes and relative phase shifts of the pressure and density
perturbations are given by:
\begin{equation}
|R|=\sqrt{ \frac{1}{(2Hk_z)^2}+1}
\end{equation}
\begin{equation}
|P|=\sqrt{ \frac{1}{(2Hk_z)^2}\left(\frac{\gamma-2}{\gamma}
\right)^2+1}
\end{equation}
\begin{equation}
\phi_R=\arctan\left(\frac{1}{2Hk_z}\right)
\end{equation}
\begin{equation}
\phi_P=\arctan\left(\frac{\gamma-2}{\gamma}\frac{1}{2Hk_z}\right)
\end{equation}

Given the wave frequency $\omega$, the vertical wave vector $k_z$
is found from the usual dispersion relation for acoustic-gravity
waves $k_z=\sqrt{(\omega^2-\omega_c^2)}/c_S$. The wave vector
$k_z$ is zero if the driving frequency is below the local cut-off
frequency.


As the atmosphere at the bottom boundary is not isothermal, the
chosen form of driver is not an exact solution for this layer. The
vertical dependence in equations (\ref{eq:v}--\ref{eq:p}) would be
slightly different including the temperature gradient. We have
checked the implications of this form of driving on the solution
developed in the physical domain. We found that varying
artificially the vertical dependence in the equations
(\ref{eq:v}--\ref{eq:p}) till the extreme case of oscillations in
the non-stratified atmosphere has negligible influence on the
solution in the physical domain. This gives us confidence that the
numerical solution is not biased by this particular form of
driving.

\begin{figure*}
\center
\includegraphics[width=17cm]{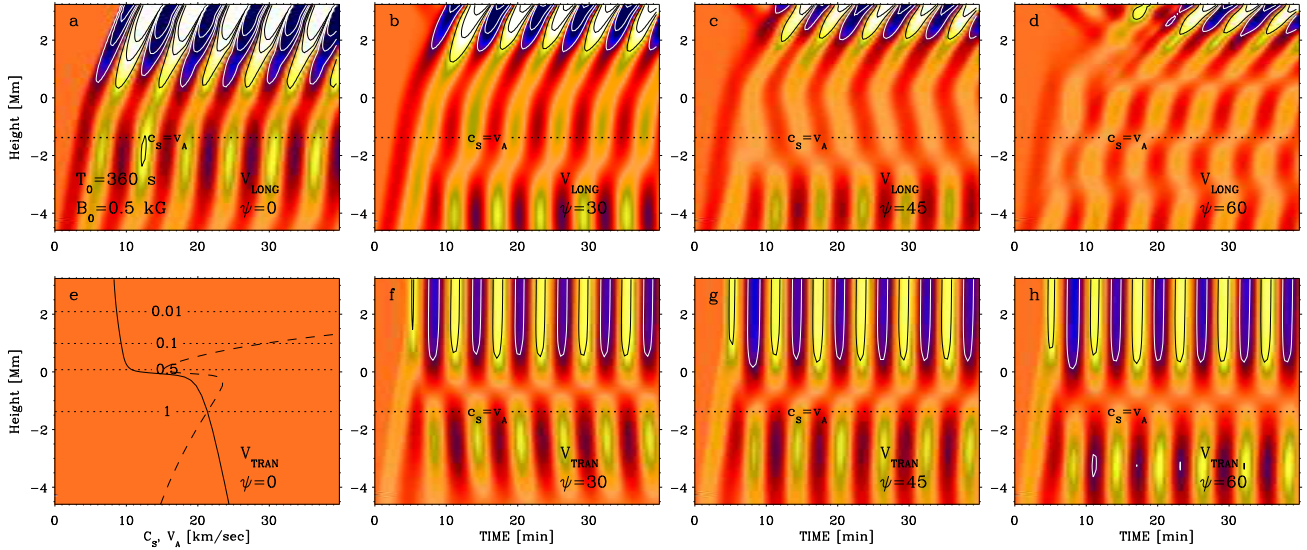}
\caption{Height-time variations of the longitudinal (parallel to
the field, top)  and transversal (perpendicular to the field,
bottom) velocities for $B_0=$ 0.5 kG, $T_0=$ 360~s and
inclinations equal to $\psi=$ 0, 30, 45 and 60$^\circ$. Yellow and
white colors mean positive velocities and dark red and blue colors
mean negative velocities. The velocity contours of $|V|= $ (1,
0.5, 0.3) \kms\ are overplotted as solid lines. Zero height
corresponds to the photospheric base ($\log\tau_5=0$). Dotted
lines marked with numbers are contours of constant $c_S^2/v_A^2$.
Height dependences of the sound speed $c_S$ (solid line) and
Alfv\'en speed $v_A$ (dashed line) are plotted over the left
bottom panel.}\label{fig:wave1}
\end{figure*}

\begin{figure*}
\center
\includegraphics[width=17cm]{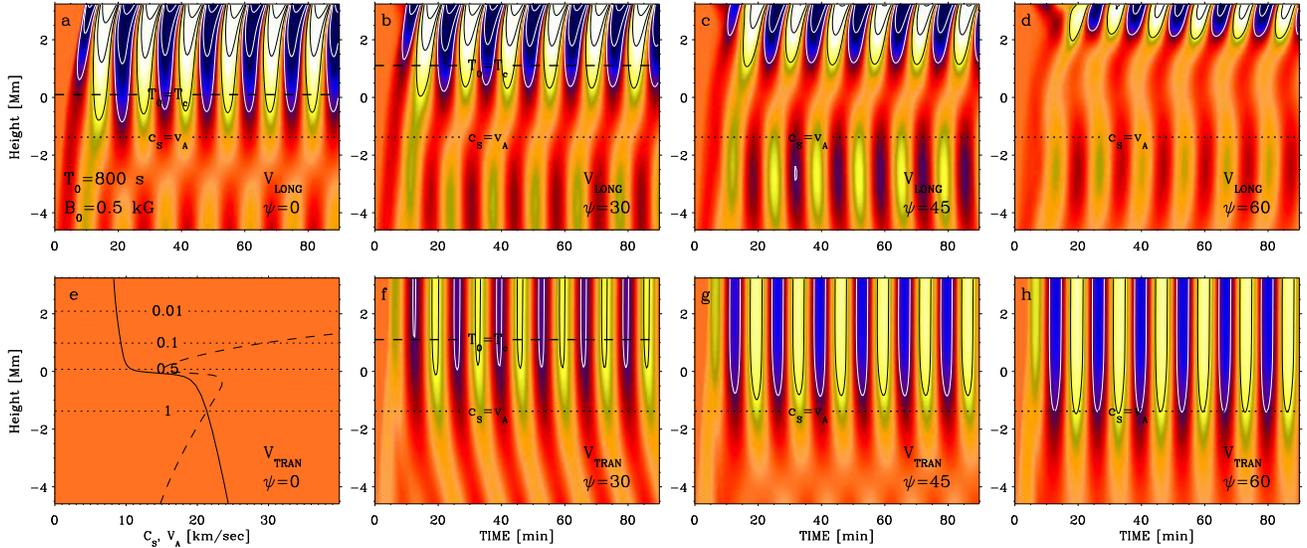}
\caption{Same as Fig.~\ref{fig:wave1} but for $T_0=$ 800~s. The
dashed line marks the location where the wave period is equal to
the cut-off period $T_0=T_c/\cos{\psi}$. }\label{fig:wave2}
\end{figure*}

\begin{figure*}
\center
\includegraphics[width=17cm]{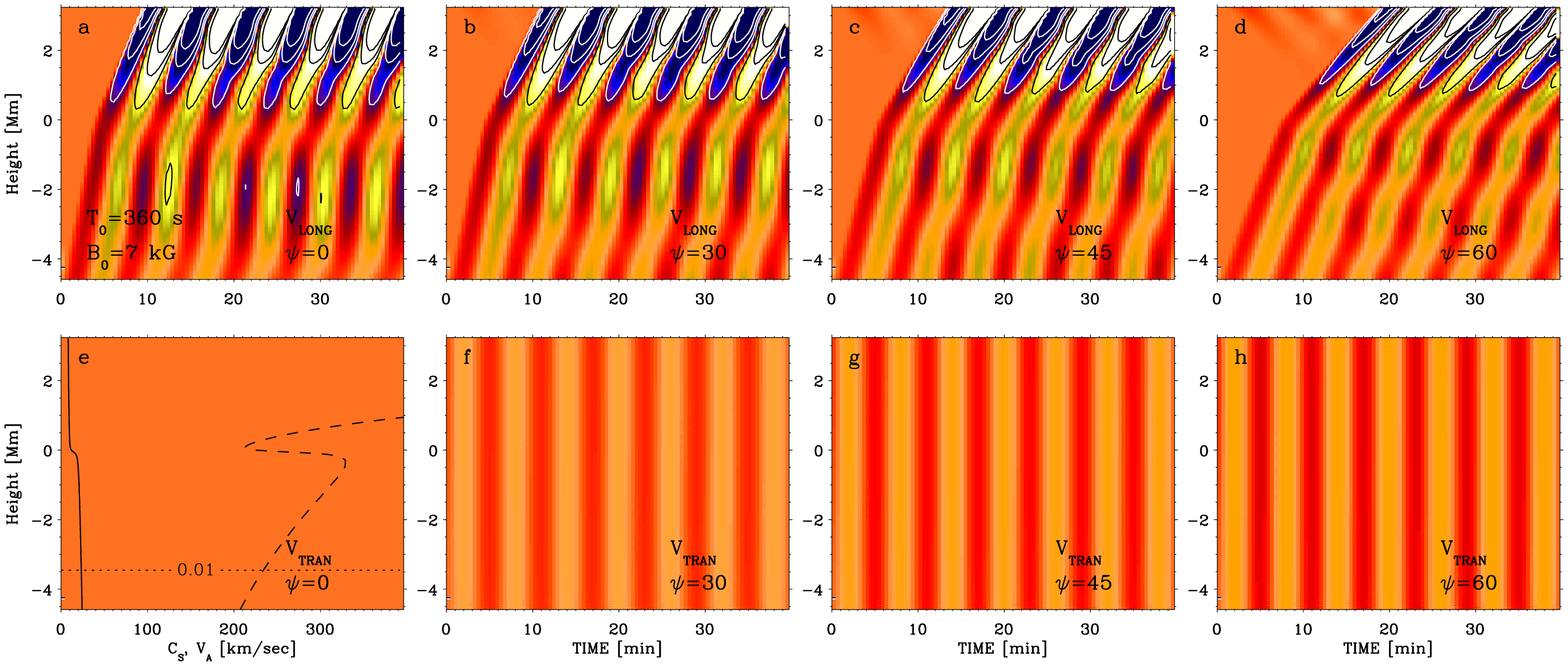}
\caption{Same as Fig.~\ref{fig:wave1} and Fig.~\ref{fig:wave2} but
for $B_0=7$ kG and $T_0=$ 360~s. }\label{fig:wave3}
\end{figure*}

\begin{figure*}
\center
\includegraphics[width=17cm]{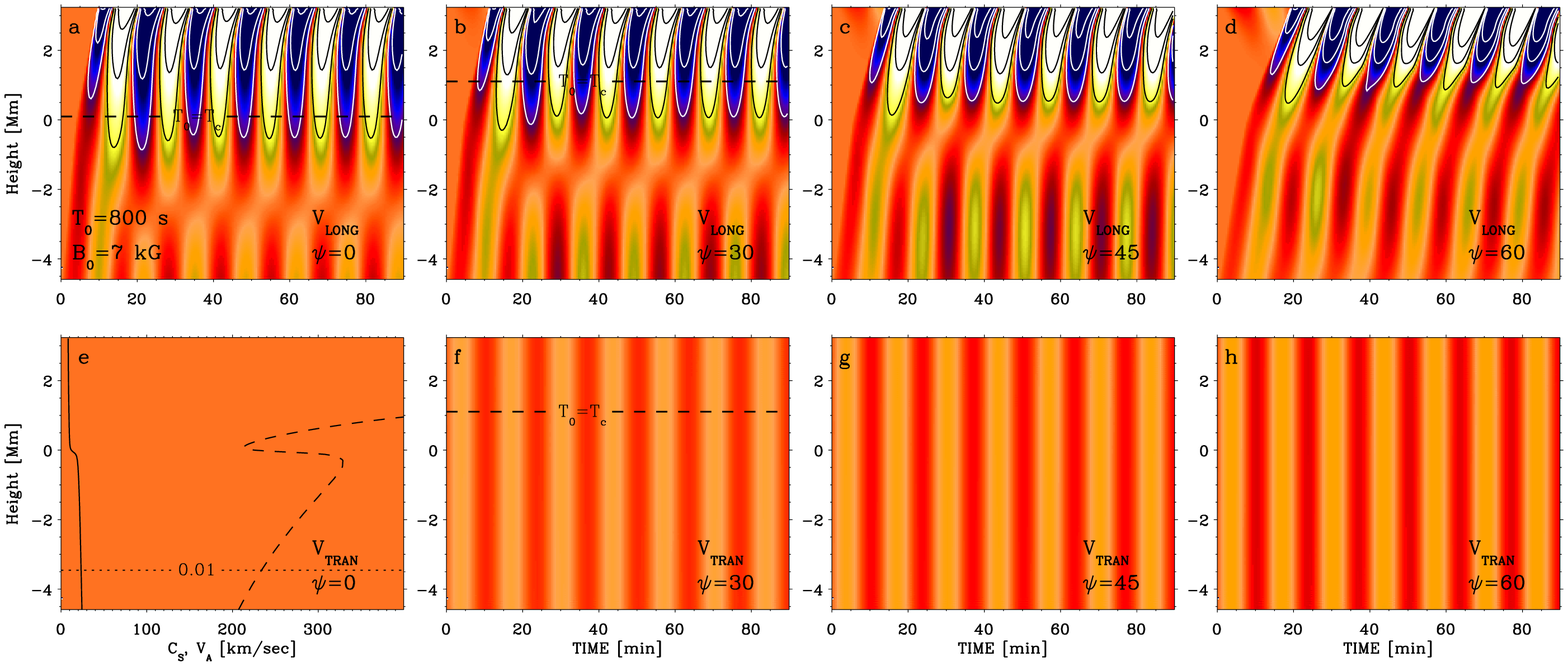}
\caption{Same as Fig.~\ref{fig:wave1} and Fig.~\ref{fig:wave2} but
for $B_0=7$ kG and $T_0=$ 800~s. }\label{fig:wave4}
\end{figure*}

\subsection{Model atmosphere and simulation setup}
\label{sect:setup}

As unperturbed equilibrium model atmosphere we adopt a model
computed with the help of the opacity-sampling model atmosphere
code LLmodels \citep{shulyak:2004}. This plane-parallel, LTE model
has effective temperature $T_{\rm eff}$\,=\,7750~K and
gravitational acceleration $\log g$\,=\,4.0 (cgs units). The model
takes into account abundance anomalies typical of cool magnetic Ap
stars and is part of the grid presented by
\citet{kochukhov:2008d}. The dipolar magnetic field is force-free
and, therefore, does not influence the hydrostatic equilibrium of
the model.

Some parameters of the model are displayed in
Fig.~\ref{fig:model}. It gives the temperature, density,
characteristic speeds and cut-of period stratifications as a
function of geometrical height and continuum optical depth
$\log\tau_5$. Several features of the model should be emphasized
for better understanding the simulations. The temperature has a
very strong gradient at the surface decreasing almost a factor of
three in 0.5 Mm. This has consequences on the critical period
$T_c$ and on the acoustic speed $c_S$, which both fall abruptly at
the atmospheric base. The density stratification has a bump around
zero kilometers ($\log\tau_5=0$), which appears due to the
hydrogen ionization in these layers. Due to this bump and due to
the temperature gradient, we can expect strong reflection of waves
coming from the interior to the surface. As a consequence of the
density stratification, the Alfv\'en speed (Fig.~\ref{fig:model}c)
has a complex behavior showing a local minimum around zero
kilometers.

In the present paper we perform a grid of simulations made for
different combinations of three parameters: local magnetic field
strength, its inclination and driving period. The magnetic field
strength grid has 5 values: $B_0=$ 0.5, 1, 2, 3 and 7 kG. The
inclination grid has 4 values: $\psi=$ 0, 30, 45 and 60 degrees to
the local vertical. And the period grid has 3 values: 360, 600 and
800 s. This makes in total 60 models. In this way we span the
typical range of magnetic field strengths and pulsation periods
observed in roAp stars. The initial amplitude of oscillations at
the base of the simulation domain was $V_0=100$ \ms\ for all
simulations. Through this paper we take positive sign for velocity
corresponding to a redshift (downflow).


Note that we do not introduce any latitudinal dependence of the
driving velocity. We use the same form of the driver
(Eq.~\ref{eq:v}--\ref{eq:p}) for all inclinations of the magnetic
field, i.e. at all latitudes of the star. In other words, we study
the response of the stellar atmosphere to an acoustic wave
perturbation introduced from below. By doing so we neglect the
possible presence of other wave types at heights of the lower
boundary of our simulation domain. For the range of the considered
magnetic field strength, the low boundary is the region where the
asymptotic solutions for different decoupled modes still can not
be safely applied. The weight of the different modes at these
heights depends on their propagation, reflection and
transformation in the deeper layers and is apriory unknown.
Several theoretical studies \citep[\eg\ ][]{Sousa+Cunha2008} show
that the presence of magnetic boundary layer has strong
implications on the latitudinal dependence of the amplitudes and
phases of waves at the bottom of the atmosphere. However, we can
not use the theoretical eigenfunctions calculated in such studies
directly in application to our simulations, as the atmospheric
model considered in them is different from ours. Giving these
limitations, the reader has to keep in mind that the results of
our calculations, as for the relative amplitude and phase
distributions of waves at different latitudes
(Sect.~\ref{sect:compare}), are strictly valid only assuming
acoustic waves at the bottom of the atmosphere with the
latitude-independent properties.

Fig.~\ref{fig:model}c gives the Alv\'en speed for the $B_0$ grid
values. As we see from the figure, already for a field strength as
weak as 1 kG, the Alfv\'en speed exceeds the acoustic speed in the
whole height range, making the atmosphere magnetically dominated.
Only for $B_0=0.5$ kG the plasma is pressure dominated below $-1$
Mm. Thus, we expect to see some mode transformation in the
corresponding simulations.

In the magnetically dominated atmosphere, the cut-off frequency
changes with the magnetic field inclination angle as the acoustic
wave propagation becomes field-aligned. Fig.~\ref{fig:model}d
illustrates acoustic cut-off periods for the inclination grid of
our simulations. This figure also shows the driving periods $T_0$.
In our model atmosphere waves with $T_0=360$ s are propagating at
all heights. However, waves with $T_0=800$ s are propagating only
in the interior and become evanescent at the surface. Thus, some
reflection of the 800~s-period waves is expected at the surface.


Through the rest of the paper, we describe the results of our
simulations in terms of fast and slow magneto-acoustic-gravity
waves (note that we use a more concise notation, i.e.
``magneto-acoustic''). These waves are the two possible solutions
of the MHD equations (\ref{eq:den}--\ref{eq:ind}), one with a
larger propagation speed and one with a lower propagation speed.
In an academic case of a homogeneous atmosphere that can be
described by single values of Alfv\'en speed $v_A$ and the sound
speed $c_S$, the waves maintain their fast or slow properties
globally in the whole domain. As we consider the case of
non-homogeneous vertically stratified atmosphere, the propagation
speeds of the two solutions change with height. In the regions in
the atmosphere where $v_A \ll c_S$ or $v_A \gg c_S$ the two
solutions are expected to recover their asymptotic behaviour
identified with classical slow and fast waves of the homogeneous
atmosphere. In the intermediate situation when $v_A < c_S$, one
can still identify the fast mode as mostly acoustic and a slow
mode as mostly magnetic. The opposite is true in the region where
$v_A > c_S$ \citep[see, e.g.][where such a classification is
applied]{Cally2001, Bogdan+etal2003, Khomenko+Collados2006}. The
modes can be considered as ``coupled'' in the region where $v_A
\approx c_S$. The later term means that their properties become
close. Even in this situation we preserve the notation ``fast`''
and ``slow'', as the propagation speed of one solution is always
larger than the other one, except when $v_A=c_S$ strictly.

\section{Results}

\subsection{Velocity variations}


According to Fig.~\ref{fig:model}, the simulated part of the
atmosphere is in the magnetically dominated regime down to field
strength of 1 kG. In this regime the natural reference system for
oscillations is the one defined by the magnetic field. Below in
this section we discuss oscillations of velocity parallel to the
field and perpendicular to the field ($V_{\rm long}$ and $V_{\rm
tran}$, respectively). This way the two, fast and slow,
magneto-acoustic modes can be distinguished
\citep{Khomenko+Collados2006}.


As mentioned above, the clear distinction between the two modes
can be done only in the situation where $v_A \gg c_S$ or $v_A \ll
c_S$. In the  $v_A \gg c_S$ limit, the slow mode is mostly an
acoustic-gravity wave guided along the magnetic field lines and is
visible in the longitudinal velocity and in the gas pressure
variations. The fast mode is magnetic and is visible in the
transversal velocity as well as in the variations of the magnetic
field vector.
In the $v_A \ll c_S$ limit, the slow mode (now magnetic) is better
visible in the transverse velocity variations and in the magnetic
field vector variations. The slow waves in this region are
quasi-Alfv\'enic in character \citep{Bogdan+etal2003}, thus, the
principal motion is transverse to the magnetic field and the
propagation direction is along the magnetic field. While the fast
mode in the $v_A \ll c_S$ limit is mostly acoustic and has no
relation to the magnetic field. It can have both longitudinal and
transverse velocity components accompanied by pressure
fluctuations. In the intermediate situation ($v_A < c_S$ or $v_A <
c_S$) the modes can not be so clearly distinguished, but still
preserve their mostly acoustic or mostly magnetic behaviour,
depending on the region. Note that we keep the discussion in terms
of $V_{\rm long}$ and $V_{\rm tran}$ projections even in this
intermediate situation since it is convenient to keep the same
notation over the whole atmosphere. It allows the analysis of the
different mode behavior to be done in a clear way.

Figs.~\ref{fig:wave1}--\ref{fig:wave4} show $V_{\rm long}$ and
$V_{\rm tran}$ velocities as a function of time and vertical
coordinate. We take for illustration four representative cases of
simulations. In the first case shown in Fig.~\ref{fig:wave1}, the
magnetic field strength is weak ($B_0=0.5$ kG) and the wave period
($T_0=360$ s) is well below the acoustic cut-off period. In this
simulation, half way though the atmosphere the high frequency
waves meet the layer where $c_S=v_A$ and we expect to observe wave
mode transformation and reflections. In the second case, shown in
Fig.~\ref{fig:wave2}, the magnetic field is similarly weak, but
the wave period is above the acoustic cut-off period at the upper
part of the atmosphere ($T_0=800$ s), giving the possibility to
study the behavior of the evanescent waves. The third case
(Fig.~\ref{fig:wave3}) illustrates the propagation of the
high-frequency waves ($T_0=360$ s) in a strong magnetic field
($B_0=7$ kG). Finally, the last case (Fig.~\ref{fig:wave4}) is for
low-frequency evanescent waves ($T_0=800$ s) in the strong
magnetic field. In all the cases we show simulations at four
magnetic latitude locations where the inclination of the magnetic
field to the local vertical is $\psi=$ 0 (magnetic pole), 30, 45
and 60$^\circ$. Note that the format of these figures is such that
larger inclination of the ridges means lower propagation speeds of
waves and that vertical ridges mean infinite propagation speed,
\ie\ stationary waves.

\subsubsection{High-frequency waves in the weak field}

\begin{figure*}
\center
\includegraphics[width=14cm]{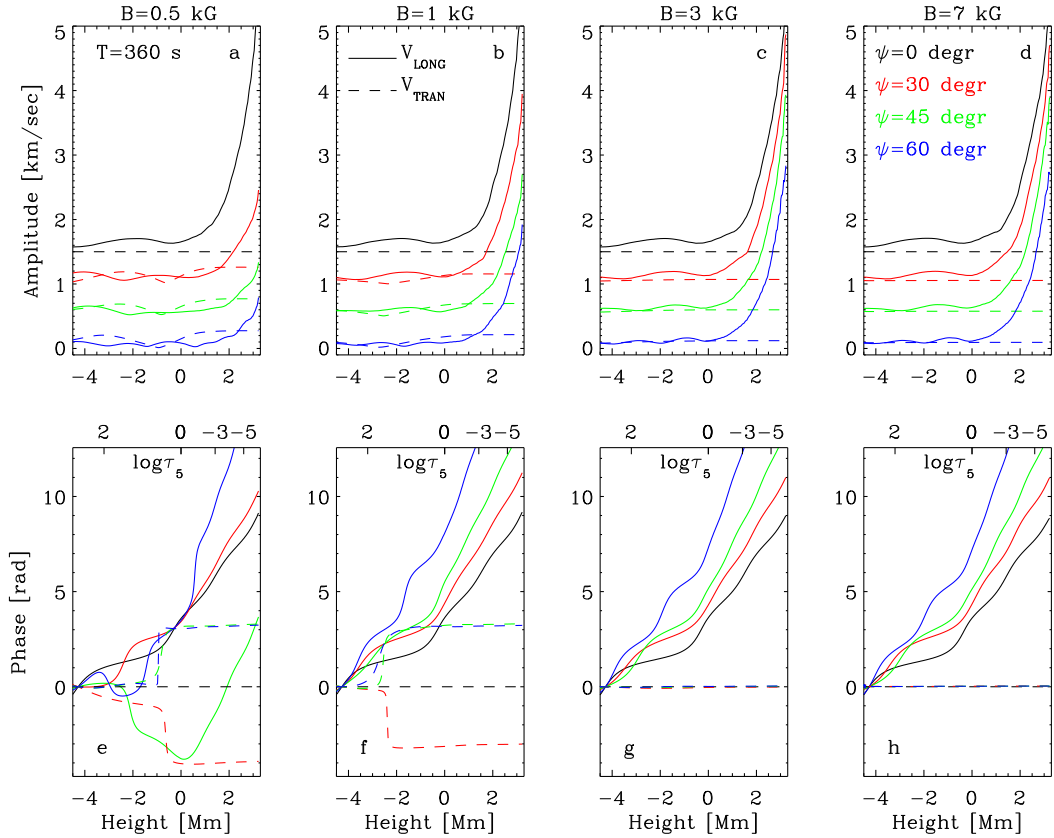}
\caption{Amplitudes (top) and phases (bottom) of the longitudinal
(solid lines) and transversal (dashed lines) velocities as a
function of height for simulations with $B_0$ = 0.5, 1, 3 and 7 kG
and at different inclinations $\psi$ (indicated in the figure).
The results are for the oscillation period $T_0=360$~s. The
amplitude curves for each $\psi$ are separated by 0.5 km/sec for
better visualization. The function fitted to the simulations is
$A\sin(\omega t+\varphi)$, so that larger phase $\varphi$
indicates a later time of pulsation maximum. }\label{fig:af360}
\end{figure*}

Fig.~\ref{fig:wave1}a shows that, as expected, in the vertical
field ($\psi=0^\circ$) only one type of wave exists with the
velocity aligned to the field. The vertical driver excites
vertically propagating fast magneto-acoustic mode that is
essentially acoustic at the bottom of the simulation domain where
$c_S > v_A$. Propagating upwards this wave encounters the layer
where $c_S=v_A$ and is completely transformed into the slow
magneto-acoustic wave (also acoustic). This mode only has
longitudinal velocity component. It can be appreciated how the
propagation speed of waves decreases above 0 km, in agreement with
the rapid decrease of the acoustic speed (Fig.~\ref{fig:wave1}e).
Fig.~\ref{fig:af360}a and e show the amplitudes and phases of the
$V_{\rm long}$ and $V_{\rm tran}$ velocities for the simulations
with $B_0=0.5$ kG and $T_0=360$~s. The amplitude of the
longitudinal velocity in the $\psi=0^\circ$ case (black lines)
does not increase monotonically, but has two local minima and a
local maximum at height around $-2$ Mm. This structure is formed
by partial reflection of the acoustic wave because of the strong
density, temperature and, thus, sound speed, gradient at $Z=0$ km.
As the frequency of the wave is above the cut-off frequency in the
whole domain (see Fig.~\ref{fig:model}) this reflection can not be
due to the cut-off frequency effects. An instructive discussion on
the partial reflection of high-frequency (above the cut-off) waves
in the atmosphere of Sun and stars is presented in
\citet{Balmforth+Gough1990}. In this paper the authors consider,
among others, the case of a polytropic interior matched to an
isothermal atmosphere with a discontinuous jump in the sound
speed. They come to the conclusion that reflection of
high-frequency waves in such a jump can be significant. Their
schematic model gives a very good approximation to our model
atmosphere where the jump in the sound speed is produced at the
photospheric base due to the hydrogen ionization. In the solar
case, partial reflections of waves from the photospheric
temperature gradient were also studied by \eg\
\citet{Marmolino+Severino+Deubner+Fleck1993}. After the reflection
the waves become partially trapped between $-3$ and 0 Mm. This is
evident from their phase behavior (black line in
Fig.~\ref{fig:af360}e) that is almost constant at these heights.
Below and above the trapped zone, the acoustic waves are
propagating upwards and their phase is monotonically increasing
with height.

The case of the inclined field is substantially more complex. The
vertical driver in the inclined field excites a mixture of the
fast and slow magneto-acoustic modes and their relative
contribution depends on the field inclination. On their way up the
fast and slow modes encounter first the transformation $c_S=v_A$
layer. In the case of the inclined field the fast-to-slow (and
vice versa) mode transformation is not complete and depends on the
wave frequency and the angle between the wave vector and the
magnetic field vector. According to \citet{Cally2006} the higher
the frequency, the smaller is the cone of angles where the
fast-to-slow mode transformation is effective.

The longitudinal velocity variations show that the fast mode
generated by the driver in the $c_S>v_A$ region is partially
transformed into the slow mode in the $c_S<v_A$ region
(Fig.~\ref{fig:wave1}b, c and d).
The comparison of the velocity amplitudes in Fig.~\ref{fig:af360}
for the case of three different inclinations $\psi$ indicates that
the part of the driver energy that finally reaches the upper layer
in the form of the slow mode might decrease with the inclination
angle.
This happens because fast-to-slow mode transformation at the
$c_S=v_A$ layer becomes less and less effective with increasing
magnetic field inclination, in agreement with \citet{Cally2006}.
Note, that, in principle, this general trend might be modified
considering more periods in the simulations. As shown by
\citet{Cunha+Gough2000}, the efficiency of the mode coupling and
its relation to the inclination of the magnetic field depends on
frequency in a highly non-linear manner.

The transversal velocity variations in Fig.~\ref{fig:wave1}f, g
and h show that the slow (magnetic) mode is produced by the driver
in the $c_S>v_A$ region. Preceding the results from
Sect.\ref{sect:therm}, the presence of the slow magnetic mode is
not only deduced from the transversal velocity variations but also
from the significant magnetic field variations present in this
region (see Fig.~\ref{fig:pb600} for $T_0=600$ sec, the results
are similar for $T_0=360$ sec). The slow magnetic mode is
partially transformed into the fast magnetic mode in the $c_S<v_A$
atmosphere. Due to the drastic increase of the Alfv\'en velocity
in the upper layers the propagation speed of the fast mode becomes
very large. The fast mode is partially reflected from such strong
Alfv\'en speed gradient region and passes again through the
transformation region. This generates additional fast and slow
modes that interfere with the upcoming waves. The downward slow
mode propagation below $-0.5$ Mm is clearly seen in
Fig.~\ref{fig:wave1}f and Fig.~\ref{fig:af360}e (red dashed line).
The node surface is formed due to the interference of the upwardly
propagating slow waves and downwardly propagating slow waves
slightly above the $c_S=v_A$ height. The exact position of this
surface depends on the vertical wavelength of the waves. Depending
on the field inclination, the mixture of the modes generated by
the driver and propagating upwards, those produced after the
multiple mode transformations and those partially reflected due to
the strong temperature gradient creates complex interference
pictures, particularly evident in the case of the largest field
inclination $\psi=60^\circ$ (Fig.~\ref{fig:wave1}d). The number
and location of the node layers depends on the field inclination,
both for $V_{\rm long}$ and $V_{\rm tran}$, and is particularly
complex in the case of weak field and low period (and, thus, low
wavelength) waves (Fig.~\ref{fig:af360}).

\begin{figure*}
\center
\includegraphics[width=14cm]{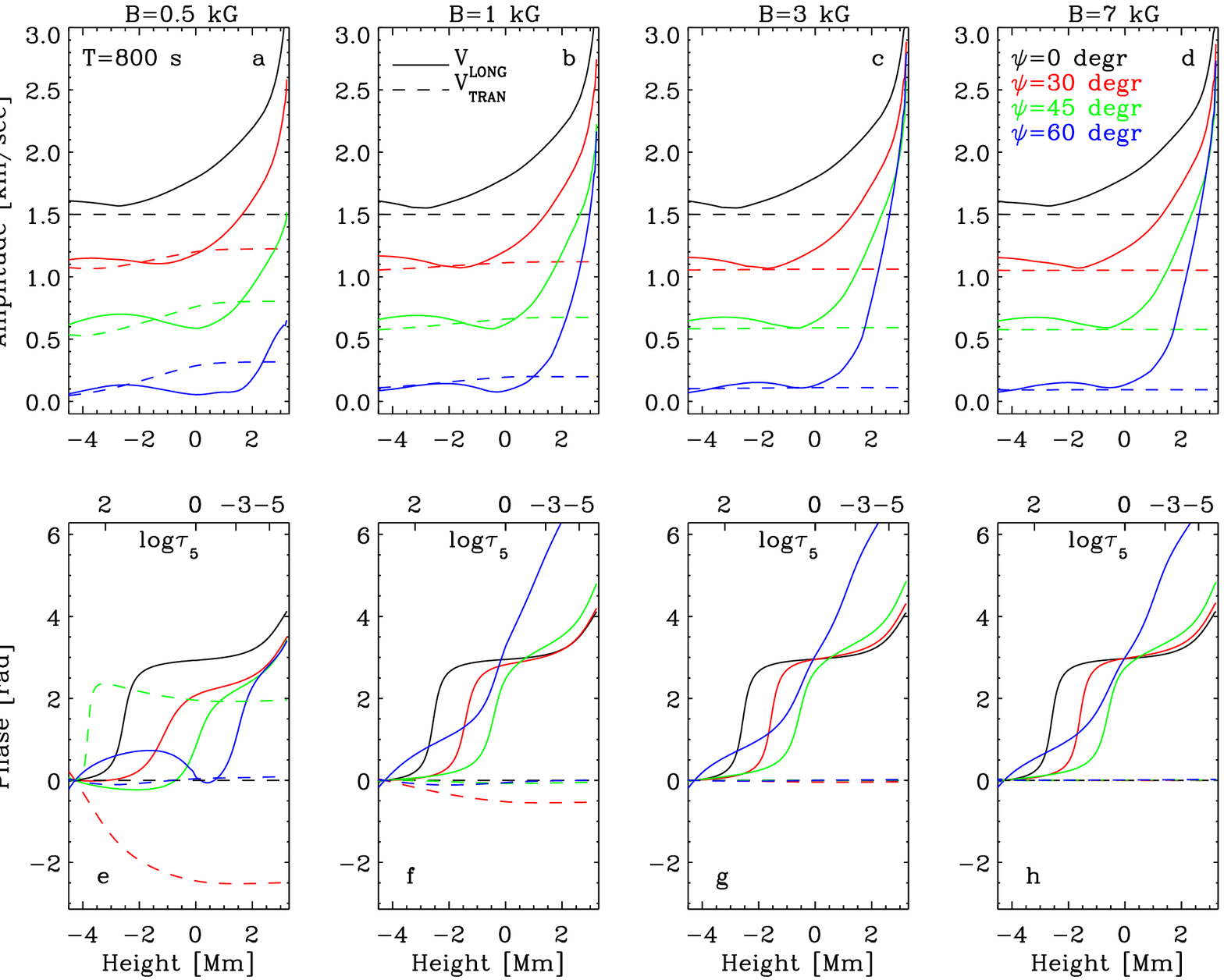}
\caption{Same as Fig.~\ref{fig:af360} but for $T_0=800$~s.
}\label{fig:af800}
\end{figure*}

\subsubsection{Low-frequency waves in the weak field}

The case of the low-frequency waves propagation in the weak field
is illustrated in Fig.~\ref{fig:wave2} and Fig.~\ref{fig:af800}
(panels a and e). In this case, apart from the transformation
layer, we have a layer  where the wave frequency becomes larger
that the cut-off atmospheric frequency. For waves with $T_0=800$~s,
this layer is located in the atmosphere where $c_S < v_A$
and, thus, its location depends on the field inclination.

In the vertical field ($\psi=0^\circ$, Fig.~\ref{fig:wave2}a) the
propagation properties are similar to the case described above.
The waves with 800~s period are reflected around $Z=0$ Mm due to
the cut-off frequency effects. Due to this reflection, a node
structure appears at about $-2.5$ Mm and a standing wave pattern
is formed by the interference of the upcoming and downward
reflected waves. Above $Z \approx 0$ Mm, the 800~s period waves
are evanescent. The evanescent wave pattern is evident from the
phase and amplitude behavior of waves in Fig.~\ref{fig:af800}a and
e. The phase is almost constant with height, except for the $\pi$
radian jump at $Z \approx -2.5$ Mm, and the amplitude increase is
much less that in the case of 360~s period waves. One may note a
propagating behavior of the waves above 2--2.5 Mm.

In the inclined field, similar to the previous case, a mixture of
the fast and slow magneto-acoustic modes is generated at the lower
boundary of the simulation domain, propagating upwards. After the
transformation layer, these modes are partially transformed and
partially transmitted. The slow magneto-acoustic modes in the $c_S
< v_A$ atmosphere are now affected by the cut-off frequency
effects. For the field inclination of $\psi=30^\circ$ the position
of the cut-off height has been shifted to $Z\approx 1$ Mm and the
waves become evanescent above this height, unlike the case of
$T_0=360$~s. With increasing inclination, the cut-off height is
moved higher up in the atmosphere outside the simulation domain
and the waves become propagating upwards. This behavior is clear
from the phase variations above $Z=1$ Mm (Fig.~\ref{fig:af800}e).
The change of the slope of the ridges in Fig.~\ref{fig:wave2}b, c
and d with increasing inclination is due to the projection effects
of the field-aligned propagation of slow mode waves.

Fig.~\ref{fig:wave2}f, g and h show how the slow magneto-acoustic
waves\footnote{Again, the slow mode in this region can be
considered as mainly magnetic as the variations in the transversal
velocity are accompanied by significant magnetic field
variations.} generated by the driver at the low boundary in the
$c_S> v_A$ region are partially transformed to the fast
magneto-acoustic waves in the $c_S < v_A$ atmosphere and again
adopt a quasi-evanescent behavior due to the rapid increase of the
Alfv\'en speed. After the partial reflection of the fast
magneto-acoustic waves,  downward propagating waves are clearly
seen below $c_A=v_A$ height. Similar to the $T_0=360$~s case,
these downward propagating waves are slow modes generated after
the secondary transformation at the $c_A=v_A$ layer. The phase and
amplitude behavior is qualitatively similar for the $T_0=360$ and
800~s waves (compare dashed curves in Fig.~\ref{fig:af360}a, e and
Fig.~\ref{fig:af800}a, e). The different pattern and node
locations are due to the different vertical wavelengths of these
waves. The height pattern is much smooth and continuous for the
waves with larger periods.

\subsubsection{High-frequency waves in the strong field}

The results of the simulations of the high-frequency wave
propagation in the strong magnetic field are illustrated in
Fig.~\ref{fig:wave3} and Fig.~\ref{fig:af360} (panels d and h).
The case of the strong field is much easier to analyze since the
wave propagation always occur in the magnetically dominated regime
and there is no mode transformation. In the high-frequency case
there are also no cut-off frequency effects and the slow
(acoustic) waves are propagating at all heights.

In the vertical field the driver generates slow magneto-acoustic
waves. In the magnetically dominated atmosphere these waves are
essentially acoustic  and their propagation properties are
identical to the case of $B_0=0.5$ kG, as the vertical magnetic
field does not affect to the first order the properties of the
vertically propagating acoustic waves \citep[see,
however,][]{Roberts2006}. In the inclined field, in the absence of
the mode transformation, the amplitudes of the slow modes that
reach the upper boundary are very similar. The propagation of
these waves is only affected by the strong gradient in the sound
speed, producing partial reflection around $Z=0$ Mm.
%
%
As the wave propagation is field-aligned, they attack the layer of
the strong sound speed gradient with a different angle. We are not
aware of any theoretical developments of the wave reflection on
the sound speed gradient in the presence of the magnetic field.
Based on the analogy with rays in geometrical optics (that applies
in the case of high-frequency short spatial wavelength waves) we
can speculate that the angle of reflection may depend on the angle
of incidence to the reflection layer. This would explain different
interference pattern formed by the upcoming and reflected slow
waves for the field with different inclinations (compare
Fig.~\ref{fig:wave3}a to d).

Fig.~\ref{fig:wave3}f, g and h show the fast magneto-acoustic
modes generated by the vertical driver. The relative contribution
of the fast mode gets larger with the field inclination $\psi$.
Unlike the case of the weak magnetic field, the phase and
amplitude of this mode remain constant with height. This happens
because the propagation speed of this mode is extremely large and
its vertical wavelength is much larger than the size of the
simulation domain.

\subsubsection{Low-frequency waves in the strong field}

Finally, Fig.~\ref{fig:wave4} and Fig.~\ref{fig:af800} (panels d
and h) illustrate the propagation of the low-frequency waves in
the strong magnetic field. This case is qualitatively similar to
the high-frequency wave propagation in the strong field, except
for the presence of the cut-off layer and a much larger vertical
wavelength of waves.

The slow magneto-acoustic waves generated by the driver are now
reflected due to the cut-off. This happens at different heights,
depending on the field inclination. As follows from the phase
dependence on height in Fig.~\ref{fig:af800}h, a standing wave
pattern is formed in the simulations with $\psi=$0, 30 and 45$^\circ$.
Unlike this, the 60$^\circ$ inclined field already decreases
the cut-off frequency sufficiently enough to allow for the
wave propagation. The phase of the slow-mode waves in the
$\psi=60^\circ$ case continuously increases with height. Note
also, that for the same reason, their relative amplitude
increase is more significant that in the case of the slow-mode
wave propagation in the less inclined magnetic field. The height
of the node layer increases with increasing field inclination.
This happens because the waves are reflected at progressively
larger heights.

The fast magneto-acoustic waves generated by the driver are
identical to the case of $T_0=360$~s.

\subsubsection{Comparative study of the amplitudes and
phases}

\label{sect:compare}

\begin{figure*}
\center
\includegraphics[width=5.7cm]{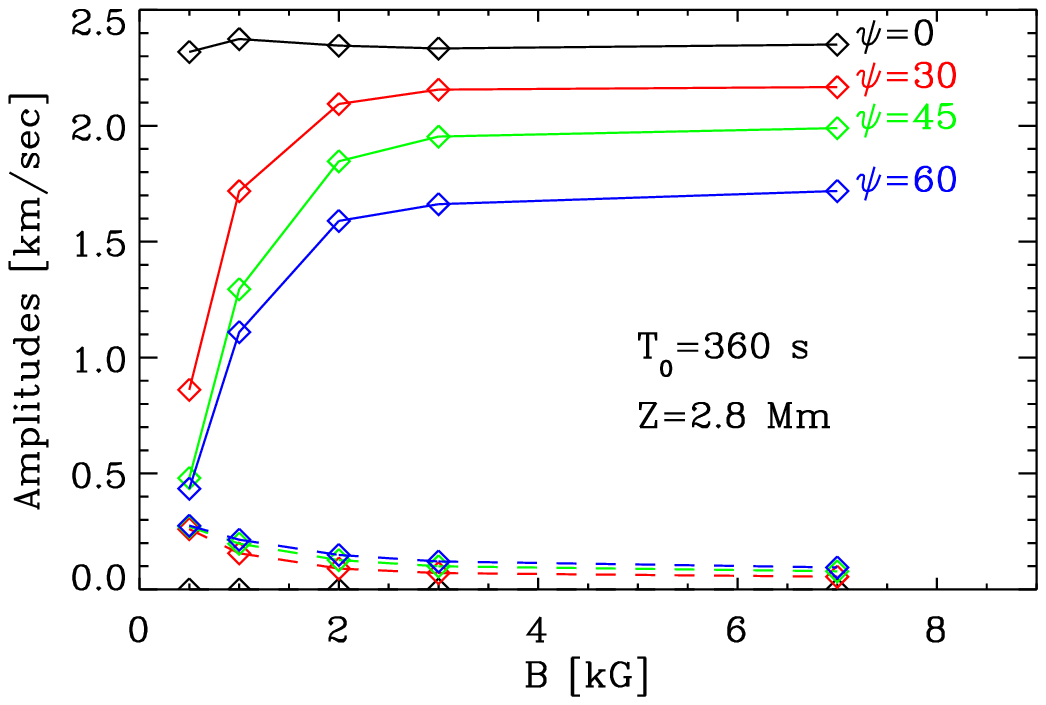}
\includegraphics[width=5.7cm]{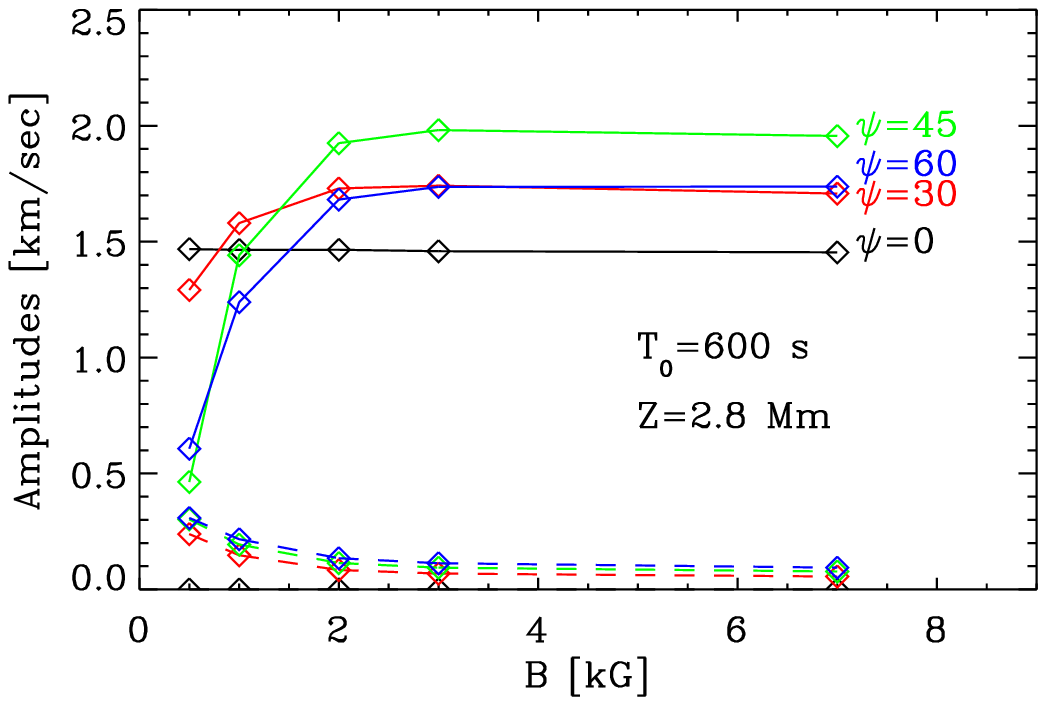}
\includegraphics[width=5.7cm]{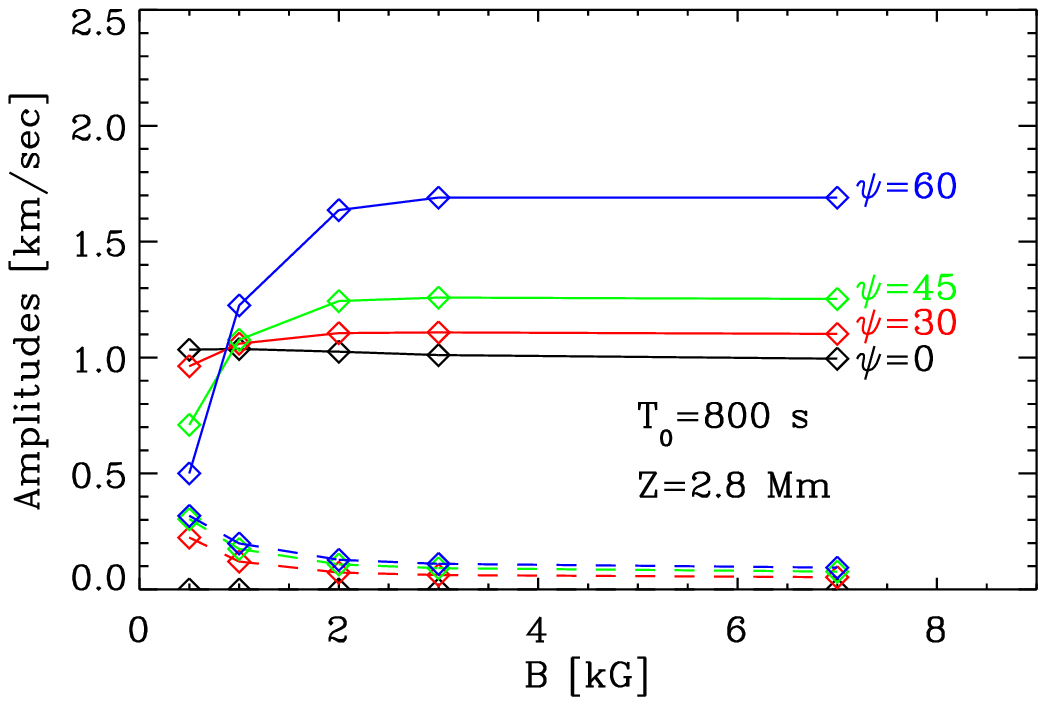}
\caption{Amplitudes measured at the top of the atmosphere at Z=2.8
Mm (corresponding to log$\tau_5=-5.2$) as a function of the field
strength for different inclinations, as indicated on the figure.
Left: period $T_0=360$~s; middle: period $T_0=600$~s; right:
period $T_0=800$~s. Solid curves: $V_{\rm long}$; dashed curves:
$V_{\rm tran}$. }\label{fig:amplitude}
\end{figure*}

Figs.~\ref{fig:af360}, \ref{fig:af800} and \ref{fig:amplitude}
show the height dependence of the velocity amplitudes and phases
for the different field strengths and inclinations. The first two
figures were partially discussed above.
%
%
Please recall that, as explained above in Sect.~\ref{sect:setup},
the comparison of the amplitudes and phases should be done keeping
in mind the particular form of the driver at the lower boundary
(i.e. acoustic wave with properties independent on the magnetic
field inclination). The relation between the amplitudes of the
fast and slow modes for a given inclination $\psi$, as well as the
relative amplitudes of waves at different $\psi$ can be modified
assuming another form of the driver.

Fig.~\ref{fig:af360} shows that for the frequencies well above the
cut-off frequency ($T_0=360$~s) the field-aligned velocity
variations have a character of the upward propagating waves for
the field strengths above 1 kG. Only for the lowest field
strengths in our simulations, $B_0=$0.5 kG, the phase behavior
with height is more complex and downward wave propagation is
observed for some angles. This complex behavior is produced by the
wave transformation and mode mixing at the $c_S=v_A$ layer. It
means that for the weakest fields the high-frequency waves can be
trapped, unlike the pure acoustic case. While the vertical
propagation is unaltered by the presence of magnetic field, the
magnetic effects become more pronounced with increasing
inclination.
For fields above 1 kG the waves propagate in the atmosphere
dominated by the magnetic field and their behavior is more simple.
The velocity component perpendicular to the magnetic field shows
the behavior characteristic of standing waves for the field
strength above 3 kG. In general, both amplitude and phase curves
get smoother and less node structure is formed with increasing
field strength.

Fig.~\ref{fig:af800} shows that the field-aligned velocity
variations for frequencies close to the cut-off frequency
($T_0=800$~s) mostly behave as evanescent trapped waves.  The
only exception is the situation of the strongly inclined field
sufficient to lower down the cut-off frequency. According to
Fig.~\ref{fig:af800}, the low-frequency waves are propagating for
$\psi=60^\circ$. It means that the low-frequency waves that
would be trapped in the absence of the magnetic field can escape
at latitudes near the magnetic equator if the field strength is
sufficiently high (above $B_0=0.5$ kG in our simulations).
Again, due to the mode transformation, the height dependence of
the amplitudes and phases is more complex in the $B_0=0.5$ kG
case. The transversal velocity variations show downward wave
propagation for the weakest fields and standing waves for the
stronger fields.

Velocity variations for the $T_0=$600~s period show an
intermediate picture between the $T_0=$800 and 360~s cases. In
the $B_0=0.5$ kG field case multiple reflections and phase mixing
are present. The cut-off period is just around 600~s at the very
top of the atmosphere (Fig.~\ref{fig:model}). Because of that the
waves in the $\psi=0^\circ$ case are reflected and a standing wave pattern
is formed. In all other cases the  waves are mostly propagating
upwards. The amplitude increase with height and the location of
the nodes is more similar to the 800~s case, as the wave frequency
is closer to the cut-off frequency and the wavelength is much
larger than in the $T_0=$360~s case.

\begin{figure}
\center
\includegraphics[width=7cm]{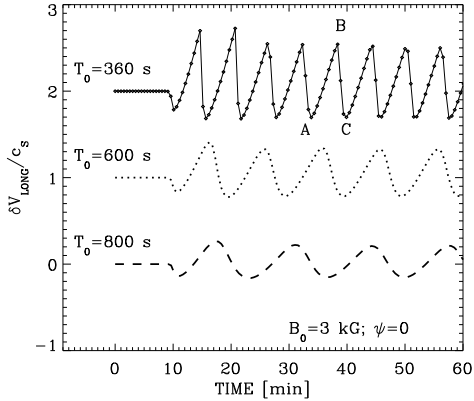}
\caption{Example of the temporal variations of the $V_{\rm long}$
at $Z=3.2$ Mm (log$\tau_5=-5.8$) for the simulations with $B_0=3$
kG, $\psi=0$ and three different periods $T_0=$360, 600 and 800 s.
The amplitudes of $V_{\rm long}$ are measured in units of the
local sound speed $c_S$. The three curves are separated by $1
\times c_S$. }\label{fig:sawtooth}
\end{figure}

\begin{figure*}
\center
\includegraphics[width=5.7cm]{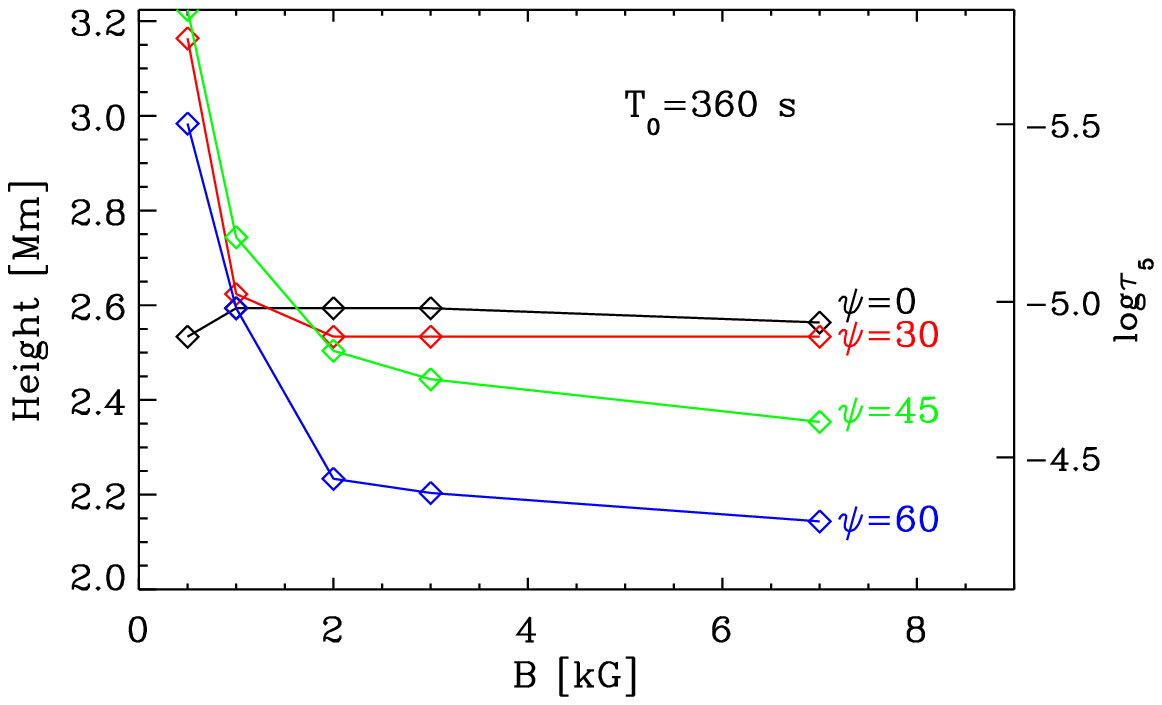}
\includegraphics[width=5.7cm]{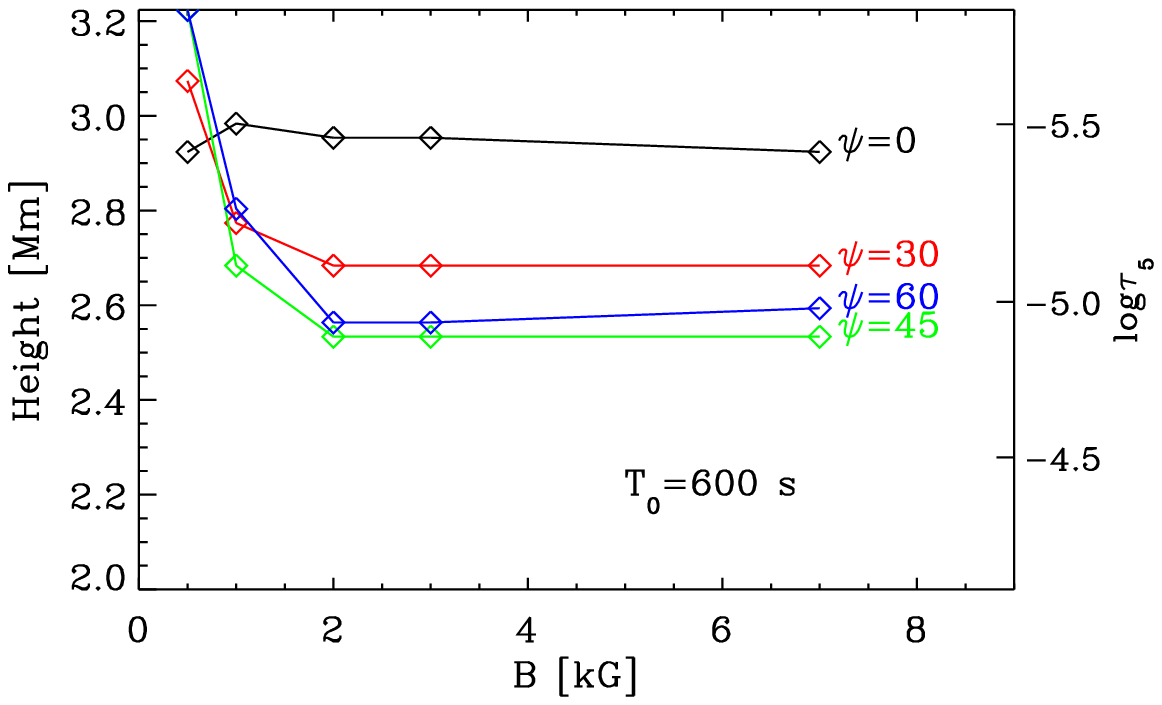}
\includegraphics[width=5.7cm]{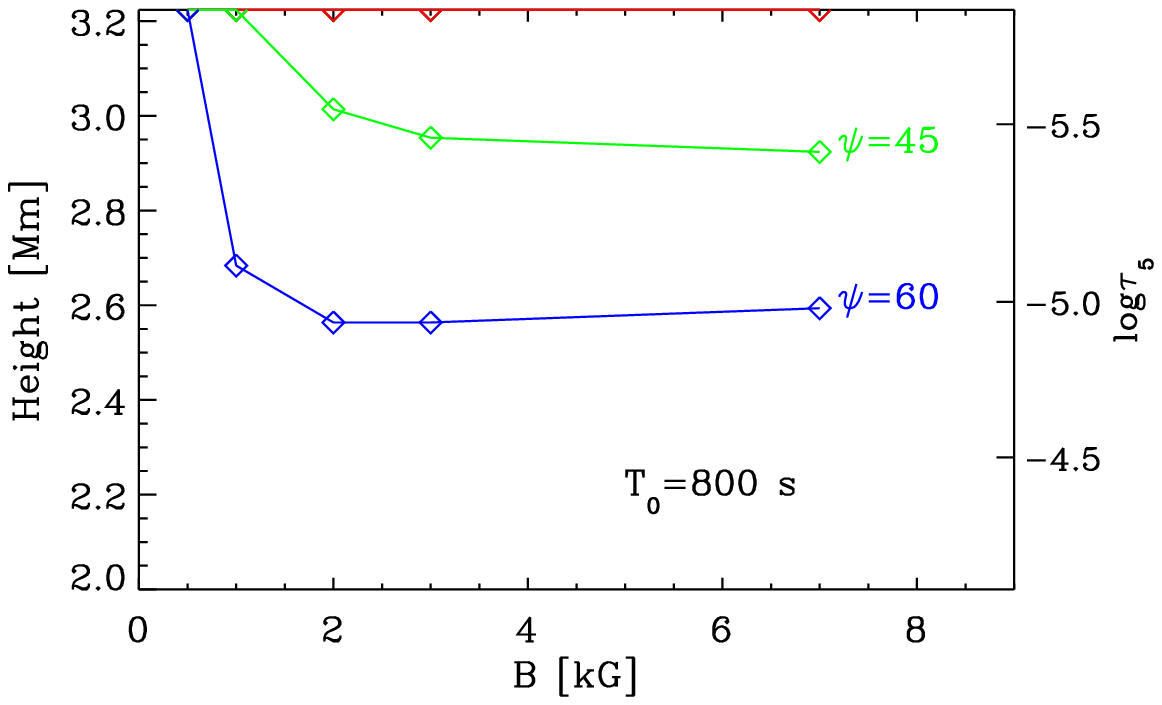}
\caption{Height of the saw-tooth wave formation in $V_{\rm long}$
as a function of the field strength for different inclinations, as
indicated on the figure) . The right vertical axis gives the
optical depth log$\tau_5$ for the corresponding geometrical
heights. Left: period $T_0=360$~s; middle: period $T_0=600$~s;
right: period $T_0=800$~s. }\label{fig:choques}
\end{figure*}


Fig.~\ref{fig:amplitude} give the comparison between the velocity
amplitudes at the top of the atmosphere as a function of the field
strength for the different inclinations and periods. It provides
information of how much of the initial driver energy reaches the
top layer and how much of it is retained below due to different
processes. First we see that for the vertical field the amplitudes
of the longitudinal velocity are field-independent. The absolute
values of the amplitudes in the vertical field case decrease with
increasing the period. Both properties are in agreement with the
analytical models, as  the vertically propagating acoustic waves
are not affected by the presence of the vertical field. Is gives
us an additional confidence in the validity of the simulations.

For the inclined field, the $V_{\rm long}$ amplitudes depend on
the field and this dependence is different for the different
periods. For the high-frequency waves ($T_0=360$ s, left panel of
Fig.~\ref{fig:amplitude}) the amplitudes decrease with the field
inclination for all $B_0$. This effect is due to
non-linear behaviour of waves. The velocity amplitude of the
linear the high-frequency waves scales as $\exp(z/2H)$ with
height. Preceding the results from Fig.~\ref{fig:choques}, the
non-linearities become important at lower heights for propagation
at larger inclination angles. The saw-tooth profile develops and
the amplitude increase with height becomes slower than
exponential. Thus, in the high-frequency case, the largest
amplitudes should be observed near the magnetic poles. This
situation is the opposite to the low-frequency wave case
($T_0=800$ s, right panel of Fig.~\ref{fig:amplitude}). The
velocity amplitude for the linear waves with frequencies below the
cut-off frequency scales with height as $\exp( z/2H
-z\sqrt{(\omega_c^2-\omega^2)}/c_S)$, making its increase with
height weaker than in the high-frequency case. However, as the
effective cut-off frequency becomes lower for the waves
propagating along the inclined field, the second term in the
exponent becomes smaller and finally disappears for sufficiently
large inclinations as the waves become propagating again. As
follows from the right panel of Fig.~\ref{fig:amplitude}, this
effect is dominating over the effect of non-linearities.  Thus, in
the low-frequency case the longitudinal velocity amplitudes
increase with $\psi$ and are the largest near the equator of the
magnetic dipole. The case of $T_0=600$ s (middle panel of
Fig.~\ref{fig:amplitude}) is the intermediate between these two.
Note that for $\psi=60^\circ$, the amplitudes are in fact the same
for all $B_0$ and $T_0$.

For all periods, the amplitudes of the longitudinal velocity at
the top of the atmosphere are the lowest for the weakest field
strength $B_0=0.5$ kG. At the same time, the amplitudes of the
transversal velocity are the highest. This is a direct effect of
the mode transformation, as  the part of the energy of the fast
mode $c_S> v_A$ waves excited at the lower boundary does not reach
the top in the form of slow mode acoustic-like waves but it is
transformed to fast mode magnetic waves observed in the
transversal component of the velocity. Note that the effects of
the mode transformation become more pronounced with increasing
inclination, as the fast-to-slow mode transformation becomes less
effective. In the $B_0=0.5$ kG case the dependence of the
longitudinal velocity amplitudes on $\psi$ is qualitatively the
same for different periods. With increasing field strength, the
effects of the mode transformation disappear and the amplitudes
become distributed according to the acoustic cut-off frequency
effects as described above. It means that, if the field is
sufficiently weak and the mode transformation can happen in the
atmosphere, the amplitudes of the longitudinal velocity are the
largest at the pole of the magnetic dipole, both for the
high-frequency and for the low-frequency waves. Note that in all
cases the transversal velocity amplitudes at the top of the
atmosphere are very small compared to the longitudinal velocity
and would contribute very little to the disc-integrated signal.

\begin{figure*}
\center
\includegraphics[width=17cm]{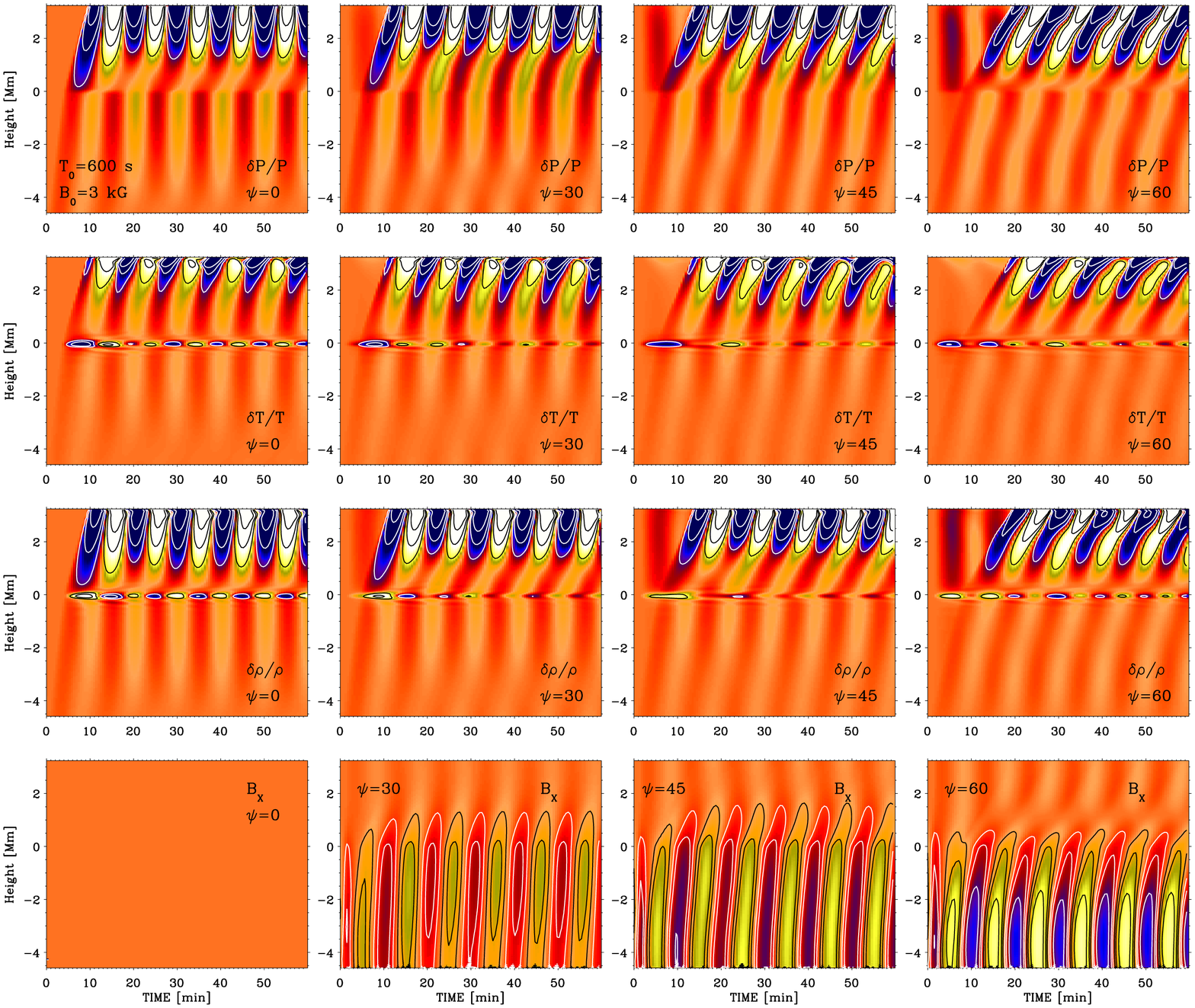}
\caption{Height-time dependences of relative variations of the gas
pressure (first row), temperature (second row), density (third
row) and horizontal component of the magnetic field vector (last
row) for the simulation with $B_0=3$ kG and $T_0=600$ s and
inclinations equal $\psi=$ 0, 30, 45 and 60$^\circ$.
Yellow and white colors mean positive values while dark red and
blue colors mean negative values.
The contours of $|\delta P/P|$, $|\delta T/T|$,
$|\delta\rho/\rho|=$ (4, 10, 20) \% are plotted as solid lines
over the first three panel rows. The contours of $\delta B_x=$
(0.2, 0.5, 1) G are plotted over the last row of panels. Zero
height corresponds to the photospheric base at
log$\tau_5$=0.}\label{fig:thermo}
\end{figure*}

\begin{figure*}
\center
\includegraphics[width=14cm]{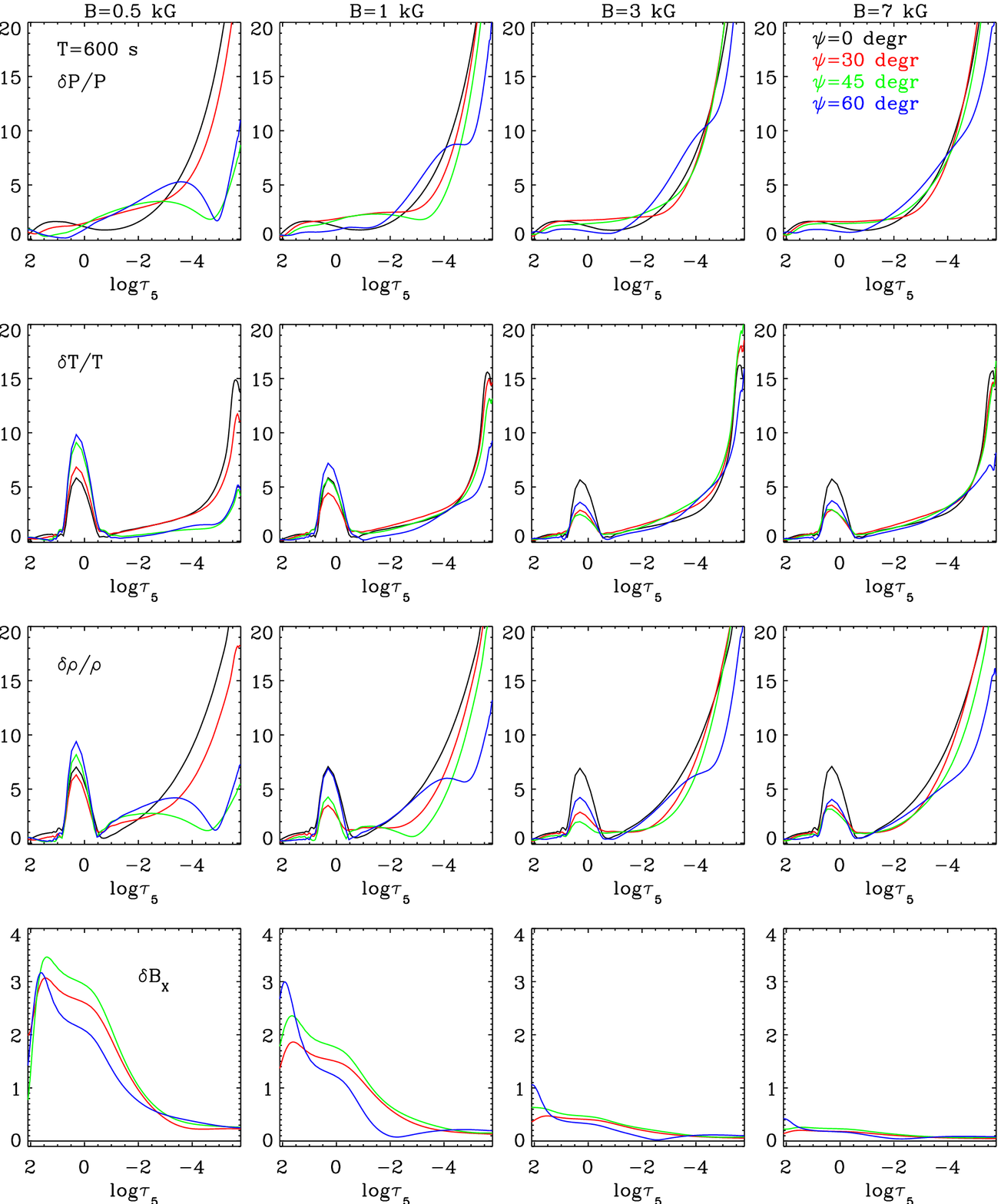}
\caption{Amplitudes of the relative variations of the gas pressure
(first row); temperature (second row); density (third row) and
horizontal component of the magnetic field vector (last row) as a
function of the optical depth log$\tau_5$. Columns from left to
right are for the magnetic field strengths $B_0=$ 0.5, 1, 3 and 7
kG. Different curves are for different inclinations, the color
coding is indicated in the figure. }\label{fig:pb600}
\end{figure*}

In summary, in our simulations, all the waves in the weak-field
case are partially trapped near the magnetic equator. In the
strong-field case, the low-frequency waves are trapped near the
poles due to the acoustic cut-off effects, but are propagating
near the magnetic equator. The high-frequency waves in the strong
field case are not trapped at all magnetic latitudes.



Finally, Figs.~\ref{fig:sawtooth} and \ref{fig:choques} describe
the non-linear behavior of waves in our simulations.
Fig.~\ref{fig:sawtooth} shows examples of the temporal variations
of the longitudinal velocity at the top of the atmosphere
log$\tau_5=-5.8$ measured in units of the local sound speed for
three different periods. The saw-tooth shape of the temporal
profile is evident for the smallest period $T_0=360$ s. The
profile is still asymmetric for $T_0=600$ s but is very close to
the normal
sine-wave
shape for $T_0=800$ s. The amplitudes vary from
around $0.25 \times c_S$ for $T_0=800$ s to $0.7 \times c_S$ for
$T_0=360$ s, but stay all the time below the local speed of sound.
According to a classical definition of a shock wave, the amplitude
of the disturbance should be above the local sound speed. Thus,
the saw-tooth waves observed in our simulations are not proper
shocks, but simple non-linear disturbances produced after the wave
has run a certain distance over the atmosphere.

Fig.~\ref{fig:choques} gives the height of the saw-tooth wave
formation as a function of the magnetic field strength for
different inclinations and periods in the simulations. This height
was defined in the way explained below. In Fig.~\ref{fig:sawtooth}
we marked three extreme points of the temporal profile $A$, $B$
and $C$. If the slope between the points $B$ and $C$ is twice
steeper than the slope between $A$ and $B$ we say there is a
saw-tooth profile. The height of the saw-tooth profile formation
depends on the wave period and its amplitude \citep{Priest1984,
Mihalas+Mihalas1984} and is typically larger for long-period
waves. In our simulations with the initial amplitude of the
velocity at the base of the atmosphere of $V_0=100$ \ms\ the
saw-tooth profile formation does not occur below 2.2 Mm
($\log\tau_5=-4.5$). Fig.~\ref{fig:choques} shows that indeed, the
height of the saw-tooth profile formation is the lowest for the
lower period waves. The saw-tooth profile formation is less
effective for the weakest field as the wave transformation effects
prevent the amplitudes from rising sufficiently to produce such
profiles. In all the cases (except $B_0=0.5$ kG), the saw-tooth
profile formation height is the lowest at high inclinations $\psi$
because the distance over the which the waves run to reach the
same height is larger for the larger inclinations.

\begin{figure}
\center
\includegraphics[width=8cm]{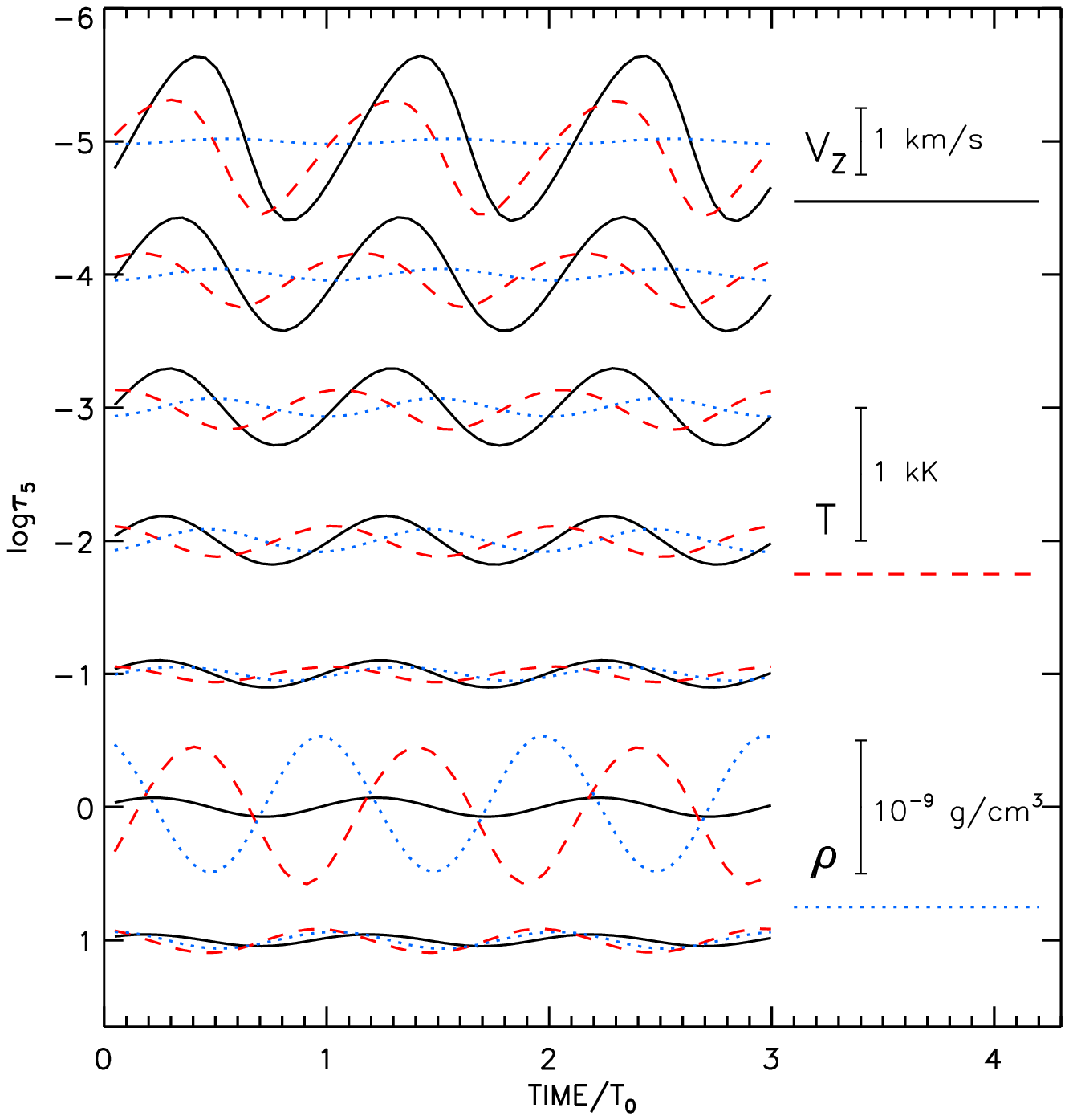}
\caption{ Absolute variations of the vertical velocity,
temperature and gas density for the simulations with $T_0=600$~s
and a vertical field with $B_0=3$~kG. Pulsation curves for
different optical depths are shifted vertically. Positive velocity
direction corresponds to redshift (downflow). }\label{fig:curves}
\end{figure}

\begin{figure*}
\center
\includegraphics[width=11cm,angle=90]{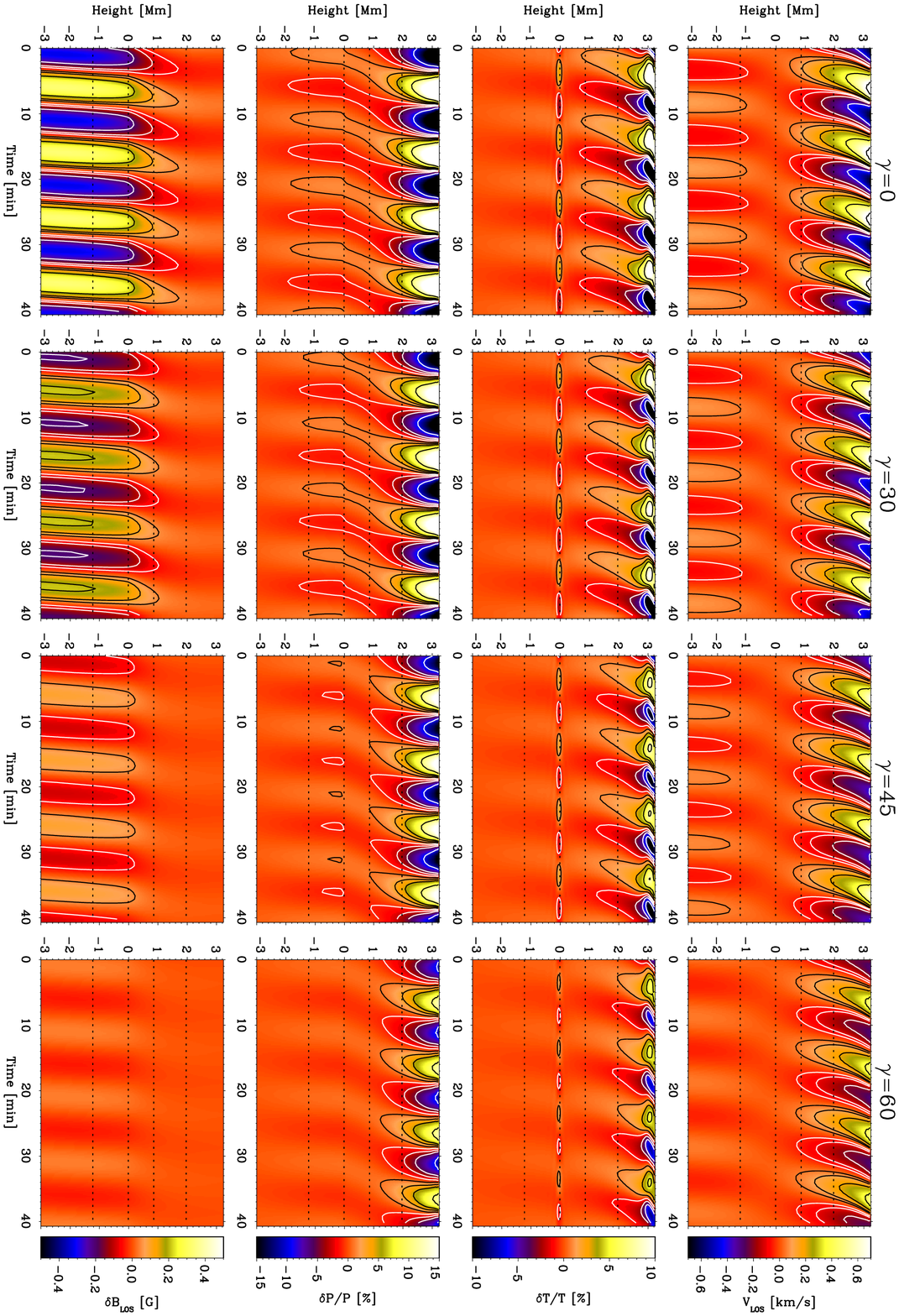}
\caption{Height-time variations of the radial velocity (first
row), relative temperature (second row), relative pressure (third
row) and the line of sight magnetic field component (bottom row)
for the global model with $B_{\rm d}=3$~kG and $T_0=600$~s.
Columns show results for different values of the angle $\gamma$
between the pulsation axis and the line of sight. The contours of
$|V_{\rm los}|=$ (0.05, 0.1, 0.2, 0.4, 0.8)~\kms, $|\delta
T/T|=|\delta P/P|=$ (1, 3, 5, 8) \%, $|B_{\rm los}|=$ (0.05, 0.1,
0.2, 0.4)~G are plotted as solid lines. The horizontal dotted
lines correspond to the optical depths $\log\tau_5=1.5$, 0, $-2$
and $-4$. }\label{fig:global}
\end{figure*}

\subsection{Variations of thermodynamic parameters and magnetic
field}

\label{sect:therm}

Our simulations allow to calculate self-consistently the
variations of the thermodynamic parameters and the magnetic field
accompanying the velocity variations. Fig.~\ref{fig:thermo} and
Fig.~\ref{fig:pb600} illustrate the pressure, temperature, density
and horizontal magnetic field variations for the simulations with
$B_0=3$ kG and $T_0=600$~s. Unlike the case of the velocity data,
we do not show simulations for the extreme cases of $B_0$ and
$T_0$ but rather discuss the intermediate situation. The behavior
of the thermodynamic variables and the magnetic field is rather
similar in all the simulations with only some quantitative
differences.

Fig.~\ref{fig:thermo} shows that the propagation of the pressure,
temperature and density disturbances from the deep layers to the
surface is not continuous. There is an amplitude and phase jump
around $Z=$0 Mm height, where the properties of the atmosphere
change in an abrupt way. An interesting feature is observed in the
density and temperature variations at this layer. There is a very
localized amplitude increase (a ``bump'') present for all field
inclinations. The nature of this localized enhancement is probably
related to the density jump located around $Z=$0 Mm.
Fig.~\ref{fig:pb600} shows that the amplitudes measured relative
to the static values are very high at the ``bump'' and can reach
some 5--10 percent. The phase of perturbations at the ``bump''
seems to suffer an abrupt 180$^\circ$ switch. The dependence of the
amplitudes at the ``bump'' on the magnetic field strength $B_0$
and inclination $\psi$ is rather complex. For the weak field
($B_0=0.5$ kG) the amplitude of both $\delta T/T$ and $\delta
\rho/\rho$ increases with the field inclination. On the other hand,
the opposite is true for larger fields ($B_0>3$ kG). Note from
Fig.~\ref{fig:pb600} that the ``bump'' is located in the region of
formation of continuum intensity around $\log\tau_5=1:-1$. Therefore,
we can expect that these strong temperature variations will lead to
continuum intensity variations which would dominate
the total flux changes of the star.
The ``bump'' is absent in the relative pressure
variations.

The amplitude of variations of the thermodynamic parameters
increases exponentially with height above log$\tau_5=-1$ and
reaches some 15--20\% at the top of the atmosphere. In all the
cases the amplitudes are smaller for larger inclinations. As
follows from the time-height pattern of the pressure and density
variations for larger $\psi$ (Fig.~\ref{fig:thermo}), at heights
above the ``bump'' both the slow acoustic-like modes and fast
magnetic modes contribute to these variations. The standing wave
pattern characteristic for the fast modes is especially visible in
the $\psi=60^\circ$ case at the first 20 min of the simulations.
Interestingly, the presence of the fast mode seems not to affect
the temperature variations. Similar to the velocity oscillations,
the oscillations of the thermodynamic parameters are more affected
by the magnetic field in the weak-field case
(Fig.~\ref{fig:pb600}, left column of panels). Several node
structures are formed in the $\delta P/P$ and $\delta \rho/\rho$
variations for $\psi=45^\circ$ and 60$^\circ$. With increasing
field strength, the height variations of the amplitudes become
very close for all field inclinations.

The last row of panels in Fig.~\ref{fig:thermo} and
Fig.~\ref{fig:pb600} give the time-height plots and the amplitudes
of the variations of the horizontal component of the magnetic
field vector  $\delta B_x$. The variations of the magnetic field
are not aware of the jump of thermodynamic parameters and their
propagation is more continuous. There are no variations of the
vertical magnetic field component. The amplitudes of variations of
the horizontal magnetic field component are  very small and do not
exceed 3--4 G. The $\delta B_x$ amplitude decreases with increasing
the magnetic field strength so that the magnetic field variations
are almost absent in the $B_0=7$ kG case. The $\delta B_x$
variations are larger in the deep layers and decrease abruptly
above log$\tau_5=-2$. The amplitudes do not have a pronounced
dependence on $\psi$. The variations are absent for $\psi=0^\circ$
as the wave propagation is purely acoustic-like.

Finally, Fig.~\ref{fig:curves} illustrates variations of the gas
density, temperature and vertical velocity component at several
different optical depths for the representative simulation set
($T_0=600$~s, $B_0=3$~kG, $\psi=0^\circ$). Unlike the previous
height-time and amplitude-height plots, this figure shows absolute
variations of the thermodynamic quantities. This format emphasizes
strong, anti-phase changes of the temperature and density at the
base of the photosphere. In the higher layers the amplitude of the
temperature variations first rapidly decreases and then gradually
increases again, reaching values comparable to the amplitude at
$\log\tau_5=0$ only in the uppermost layers. The variation of the
vertical velocity component rapidly grows above
$\log\tau_5\approx-1$ and acquires a noticeable non-sinusoidal
shape at $\log\tau_5=-5$. In these highest layers the density
variation appears to be relatively unimportant.
%
%
In the upper atmosphere, above log$\tau_5=-1$ the phase shifts
between temperature, pressure and density variations follow the
behavior characteristic for acoustic-gravity waves dominated by
buoyancy force \citep{Mihalas+Mihalas1984}. The temperature leads
downward velocity by some 0.12--0.25 of pulsation period
(45\degree--90\degree), depending on height, while the density
lags behind the velocity by nearly the same amount. At layers
around the ``bump'' at log$\tau_5=0$, the phase relations are
switched by 180\degree. Of course, details of this pulsation
picture depend somewhat on the assumed pulsation period and
magnetic field characteristics but the main features persist
through all simulations.



\section{Global model}

In addition to the extensive grid of local simulations for
different values of the pulsation period, magnetic field strength
and orientation, we performed a more restrictive calculation aimed
at reproducing the disc-integrated signal of the oblique
non-radial pulsations observed from different aspect angles. As
discussed in Sect.~\ref{sec:genprop}, the horizontal structure of
the roAp pulsation generally resembles the axisymmetric $\ell=1$
pulsation mode, aligned with the dipolar magnetic field.
Therefore, we adopted the field components:
\begin{eqnarray}
B_r & = & B_{\rm d}\cos{\theta} \\ \nonumber
B_\theta & = & 0.5 B_{\rm d}\sin{\theta} \,,
\end{eqnarray}
where $\theta$ is the latitude measured from the magnetic pole and
$B_{\rm d}$ is the polar magnetic field strength. The strength and
inclination of the magnetic field at latitude $\theta$ are given
by:
\begin{eqnarray}
B_0 & = & \sqrt{B_r^2+B_\theta^2} \\ \nonumber
\psi & = & \arccos{B_r/B_0} \,.
\end{eqnarray}

We considered two dipolar field strength values, $B_{\rm d}=0.8$
and 3 kG, and adopted the latitude-dependent driver amplitude,
$V_{\rm d}=V_0\cos{\theta}$, according to the simplest form of the
oblique pulsator model.
%
%
Apart from that we did not introduce any latitudinal dependence in
the properties of the driver as given by the equations
(\ref{eq:v}--\ref{eq:p}). As explained in Sect.~\ref{sect:setup}
the results of the disc integration are dependent on this
particular form of the driver. This should be kept in mind
analyzing the disk-integrated variations presented below.
Similar to the local models, we used $V_0=100$~\ms. The driving
period was set to 600 s. Simulations were performed for an
equidistant latitude grid in $\cos\theta$, varying from
$\cos\theta=0.95$ to 0.05. For the purpose of disk integration, we
divided each hemisphere of the star into 10 latitude zones,
corresponding to this $\theta$-grid. Each latitude was further
partitioned into 60--100 elements, giving us a total of 1148
surface zones.

Four angles are required to fully describe the orientation of the
stellar rotation and magnetic axes with respect to the observer
\citep[e.g., see Fig.~6 in][]{piskunov:2002a}. However, here we
are not interested in the transverse magnetic field component.
Moreover, we limit ourselves to the investigation of the
disk average of the line of sight velocity component (radial
velocity), which is not altered by the stationary velocity field
produced  by the stellar rotation \citep{kochukhov:2005}. With
these simplifications the only relevant parameter is the angle
$\gamma$ between the pulsation axis and the line of sight.
%
%

We performed the disk integration by transforming the surface
pulsation structure from the stellar to the observer's reference
frame, rotated the coordinate system according to the angle
$\gamma$ and summed, with an appropriate weight, different
pulsation quantities over all visible surface elements. This
procedure was applied independently for each depth layer and
simulation time step. The weight function took into account the
projected areas of the surface zones and a linear limb darkening
with the coefficient $u=0.5$. We have verified that the line of
sight magnetic field component, $B_{\rm los}$, obtained with this
numerical scheme reproduces the longitudinal field calculated with
the well-known analytical formula \citep{leroy:1994} to within
3\%.

Fig.~\ref{fig:global} presents the height-time variations of
different quantities for the global model with $B_{\rm d}=3$~kG.
In this plot a positive radial velocity $V_{\rm los}$ corresponds
to the redshift as seen by the observer. On the other hand, a
positive longitudinal field $B_{\rm los}$ points towards the
observer, according to the standard definition used in solar and
stellar magnetometry.

As expected, the amplitude of pulsations depends on $\gamma$.
However, the overall depth dependence is dominated by the
pulsations close to the magnetic pole and hence is similar for
different angles between the line of sight and the pulsation axis.
The radial velocity variations are qualitatively similar to the
behavior of $V_{\rm long}$ for the low $\psi$, strong-field,
low-frequency local models (Fig.~\ref{fig:wave4}).
%
Thus, the radial velocity pulsations are dominated by the slow
acoustic-like modes in all the visible atmospheric layers. This
result is easy to understand giving the fact that the amplitudes
of the transversal velocity variations produced by the fast
magneto-acoustic modes in our simulations are always smaller than
the amplitudes of the longitudinal velocity variations due to the
slow magneto-acoustic modes, except for the magnetic latitudes
with a very inclined magnetic field $\psi=40$\degree$-60$\degree
(see \eg\ Fig.~\ref{fig:af800}). The node structure present in the
radial velocity oscillations at log$\tau_5=0$ is the remnant of
the nodes due to the wave reflection on the density jump at the
photospheric base. This node is persistent at all latitudes and
does not disappear in the integrated velocity data. Apart from
this node, the wave amplitude in the  high layers increases
rapidly with height reaching around 1 \kms\ at the top of our
atmosphere.

Fig.~\ref{fig:global} shows that the prominent temperature
fluctuations around $\log\tau_5=0$ are retained in the
disk-integrated picture. Thus, this pulsational feature will have
an important influence on the disk-averaged observables.
%
The amplitudes of the temperature fluctuations reach a few
percent, a value similar to individual variations at each
particular magnetic latitude. The temperature variations are the
highest when the pulsation axis and the observational line of
sight are co-aligned.
The variations of the mean field modulus (not shown) exhibit a
height dependence comparable to $B_{\rm los}$ and an amplitude
below 1~G.

The global model results obtained for the dipolar field with
$B_{\rm d}=0.8$~kG are similar to those described above. The main
difference is a higher amplitude of the relative temperature
variations at the photospheric base (up to 5\%), an increase of
the magnetic field fluctuations by a factor of two and a steeper
change of the pulsation velocity phase with depth, reminiscent of
the $V_{\rm long}$ behavior in Fig.~\ref{fig:wave1}.



\section{Discussion and conclusions}

In this paper we have described the results of the simulations of
magneto-acoustic wave propagation in the atmospheres of roAp stars
for a wide grid of magnetic field strengths and wave periods. Our
simulations allowed us to obtain a more generalized picture of the
wave behavior, reflection, mode transformation, formation of
evanescent waves and node layers. Our main findings can be
summarized in a following way:
\begin{itemize}
\item The atmospheric regions above log$\tau_5=0$ of the considered
roAp star model are magnetically dominated already for the field
strengths as low as 0.5 kG. Both fast and slow magneto-acoustic
modes are present simultaneously in these layers. Excited by
vertical driving in our simulations, the slow mode waves dominate
in the regions of nearly vertical field close to the magnetic
poles. The fast mode waves only have comparable amplitudes to the
slow mode waves at latitudes close to the magnetic equator,
corresponding to $\psi$ larger that 50\degree--60\degree
(depending on the frequency).

\item The slow mode waves are mostly propagating upwards and are
field-aligned and the fast mode waves acquire a standing wave
pattern in the upper atmosphere due to their very large vertical
wavelength produced by the rapid increase of the Alfv\'en speed.
This smooth behavior is disturbed by different reflection layers:
cut-off layer, $c_S=v_A$ transformation layer and the density
``bump'' around log$\tau_5=0$. Due to reflections on (or above)
these layers, trapped waves or waves propagating down are produced
in the regions below the visible surface in our simulations.

\item The node layers can form due to the wave reflections at the
cut-off layer, at the density ``bump'' or after the fast mode
reflection above the $c_S=v_A$ layer. In addition to these
physical reasons, another nodes can be formed in the line-of-sight
velocity projection due to the interference of the fast and slow
magneto-acoustic modes. These type of nodes are more frequent at
latitudes close to magnetic equator where the amplitudes of the
modes become of the same order of magnitude.

\item In the atmosphere above log$\tau_5=0$ the amplitude of the
slow mode waves increases exponentially with height, in accordance
with behavior of acoustic-gravity waves in a stratified
atmosphere. Their phase behavior depends on the frequency and the
field inclination. The amplitude and the phase of the fast mode
waves vary only slowly with height in agreement with their nearly
standing character.

\item In the strong-field case, when the whole simulation
domain is in the field-dominated regime ($B_0> 1$ kG), the
low-frequency waves (below the cut-off) are evanescent in the
atmosphere, except for the regions close to the magnetic equator,
where the cut-off frequency is lowered due to their field-aligned
propagation. The high-frequency waves (above the cut-off) are
propagating at all magnetic latitudes in the strong-field case.
The presence of the mode transformation changes this situation. In
the weak-field case ($B_0< 0.5$ kG) both the high-frequency and
low-frequency waves become partially trapped near the magnetic
equator due to the reflections of the fast magneto-acoustic mode
produced in the magnetically-dominated part of the atmosphere.

\item The temperature and density variations show an amplitude and
phase jump around log$\tau_5=0$, where the properties of the
atmosphere change in an abrupt way. The amplitudes of temperature
and density variations are unusually high at these layers and can
reach some 5--10\% of the static values. The amplitudes of the
magnetic field variations reach at most 3--4 G, decreasing rapidly
with height and with the $B_0$ field strength of the simulation.

\item The disc-integrated variations of the line-of-sight velocity and
thermodynamic parameters mostly preserve signatures of the slow
magneto-acoustic waves close to the magnetic poles for all angles
between the line-of-sight and and pulsation axis. The amplitudes
of the variations increase exponentially with height. The node
layer in velocity is maintained around log$\tau_5=0$. The
prominent increase in the amplitudes of variations of temperature
and density around log$\tau_5=0$ is also preserved in the
disc-integrated data.

\end{itemize}

Our conclusions regarding the existence of the fast and slow
magneto-acoustic waves and their propagation properties are
similar to those obtained in the theoretical work of
\citet{Sousa+Cunha2008}. The aim of our analysis was to understand
what mechanisms produce such diverse signatures of pulsations
measured in the atmospheres of roAp stars. \citet{Sousa+Cunha2008}
addressed a different question, namely the energy losses of the
high-frequency waves associated with the presence of magnetic
field. Despite the different objectives, our modelling allows us
to make some conclusions regarding the behavior of the
high-frequency waves as well. In particular, we conclude that, if
on their way to the surfaces the high-frequency fast
magneto-acoustic waves (essentially acoustic in the interior where
$c_S>v_A$) travel through the region of equal Alfv\'en and sound
speeds where the mode transformation can occur, they can be
partially transmitted as fast magnetic waves in the magnetically
dominated atmosphere ($c_S<v_A$). The fast magnetic waves are then
reflected back to the interior and thus, some part of their energy
is lost at the surface. This way the high-frequency waves can
become partially trapped. This mechanism acts more efficiently at
high magnetic latitudes. The part of the energy of the
acoustic-like fast modes in the interior transmitted as fast
magnetic modes increases with increasing the inclination angle
between the magnetic field and the direction of the wave
propagation, thus becoming larger at the magnetic equator. At the
poles where the field is close to vertical, the high-frequency
fast acoustic-like modes from the interior are transformed
completely into the slow acoustic-like modes in the atmosphere and
no trapping occurs.

The difference between our modelling and the one by
\citet{Sousa+Cunha2008} is that we considered much smaller portion
of the stellar interior. Thus, in our simulations the wave
transformation happens only for the weakest field strengths. Due
to that, we have concluded that the high-frequency waves are
trapped near the magnetic equator only in the weakest fields,
below 0.5 kG. If we would have considered a larger portion of the
interior of the star, the mode transformation could have affected
waves for larger magnetic field strengths as well.

\begin{figure}
\center
\includegraphics[width=8cm]{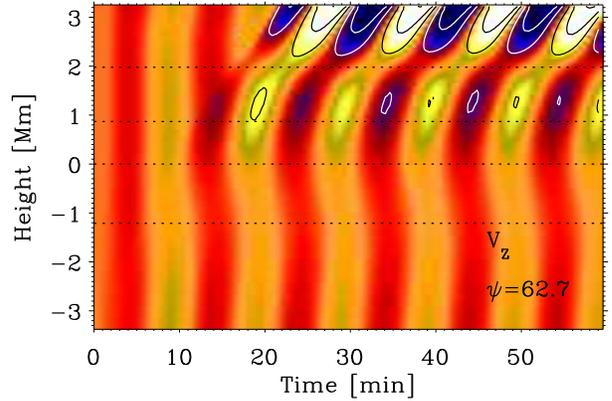}
\caption{Example of the height-time variations of the vertical
velocity for the global model with $B_{\rm d}=3$~kG and
$T_0=600$~s at the magnetic latitude corresponding to
$\psi=62.7$\degree. Yellow and white colors mean positive
velocity, while dark red and blue colors mean negative velocity.
The horizontal dotted lines correspond to the optical depths
$\log\tau_5=1.5$, 0, $-2$ and $-4$. Note the presence of a node
around log$\tau_5=-2$ due to the fast and slow mode
interference.}\label{fig:example}
\end{figure}

The low-frequency waves with frequencies below the cut-off
frequency are shown to be trapped at the magnetic poles due to the
cut-off effects. The low-frequency waves are also partially
trapped near the magnetic equator, since they are affected by the
mode transformation in a similar way as high-frequency waves.
However, it is possible that some part of their energy escapes at
intermediate magnetic latitudes where the field inclination is
already sufficient to lower down the cut-off frequency and the
fast to fast mode transmission is not complete. These conclusions
are all in line with \citet{Sousa+Cunha2008}.

Our analysis allows us to link directly the theoretically
calculated wave perturbations in all quantities to the heights in
the atmosphere, and, thus, to compare the simulated and observed
wave behavior. We stress that the conclusions coming from our
analysis remain valid, despite we take only a small portion of the
stellar interior. The relative distribution of the wave amplitudes
at different frequencies may change due to the mode
transformation, but the general picture of the propagation,
amplitude increase and the node formation will remain valid. For
example, the consequences of the wave reflection (such as the
standing wave formation or the downward wave propagation due to
the mode transformation) can hardly be detected in observations
since these processes happen below the visible surface in the our
roAp star model. Only for the weakest field strengths ($B_0 < 0.5$
kG) these features can, in principle, be detected in the deepest
observable atmospheric layers.

Our disc-integrated data suggests that, in the assumed excitation
model, the variations of all quantities in the upper atmosphere
(except for the magnetic field) are dominated by slow
magneto-acoustic modes. This conclusion is different from
\citet{Sousa+Cunha2009} who considered disc-integrated velocity
from their simplified, analytical model for the case of the
observer being pole-on in relation to the magnetic field axis.
They suggest that the observed variations are the superposition of
the acoustic running waves and magnetic standing waves and the
phase depends on the relative contribution of both components. But
for the magnetic and pulsation geometry adopted in our study
mostly the signatures of the slow acoustic-like waves remain in
the disc-integrated data.

One of the main goals of our work is to interpret the complex
observational picture of the spectroscopic variability of roAp
stars. However, detailed analysis of observations requires
computation of the hydrodynamical models and spectrum synthesis for
specific stars, taking into account their pulsation frequencies,
magnetic field strengths, orientation of pulsation and rotation
axes, and individual atmospheric structure. We plan to undertake
such an in-depth investigation of selected roAp stars in the future
studies. Here we limit ourselves to the qualitative comparison of
our model predictions with the main observational features of roAp
pulsations.

The growth of the pulsational amplitude with height, observed in
all roAp stars, is frequently considered in the context of a
simple picture where the kinetic wave energy, $\rho v^2/2$, is
approximately constant with height, so that pulsation velocity
amplitude increases due to the exponential decrease of density in
the stratified atmosphere \citep[e.g.,][]{kurtz:2007}. Following
these arguments, one can estimate an increase of velocity
amplitude by a factor of 5--10 from Fe-peak to REE lines. This is
not consistent with the observational studies, which often infer
the REE to Fe-peak velocity amplitude ratios of 30--100. Our
simulations demonstrate that the variations in the layers probed
by the lines of light and iron-peak elements are strongly
influenced by the density inversion around $\log\tau_5=0$. This
structure, located at the base of the photosphere of all dwarf
stars in the roAp temperature range, leads to the formation of the
velocity node right in the region where lines with weak or
undetectable pulsational variation are formed. This node is also
present in the disc-integrated data in the same layers, while
higher in the atmosphere the amplitude of the line-of-sight
velocity increases exponentially. Consequently, as the wave
propagates outside the node region, the velocity amplitude
increases by a factor of 30--50 in our simulations, in reasonable
agreement with observations.

The atmospheric structure around $\log\tau_5=0$ has equally
important repercussions for the pulsational temperature
variations. Our simulations predict prominent temperature changes
in this region, which will very likely dominate the observed
luminosity variation of roAp stars. Thus, our
magneto-hydrodynamical models offer a new perspective on the
photometric variability of pulsating Ap stars. We suggest that its
origin is not a smooth, weak temperature fluctuation, gradually
changing with height almost equivalently to $T_{\rm eff}$
variation \citep{medupe:1998}, but a high-amplitude,
vertically-localized phenomenon. A complete understanding of this
temperature variation is crucial for relating spectroscopic and
photometric time-resolved observations of roAp stars
\citep{ryabchikova:2007a} and for interpreting photometric
pulsations in different bands \citep{matthews:1996}. The position
and shape of the density inversion can be also an important factor
determining the visibility of the roAp pulsations in the
Hertzsprung-Russell diagram. Excitation theories
\citep[e.g.,][]{theado:2009} predict roAp-type oscillations in stars
much hotter than the empirical blue border of the roAp instability
strip at $T_{\rm eff}\approx8100$~K. It is possible that the
absence of the density inversion at the right atmospheric height
diminishes luminosity variations, making them undetectable for the
photometric pulsation surveys of hotter Ap stars.

The depth dependence of the phase of the velocity variations
obtained in our simulations shows a broad agreement with
observations. The pulsation phase increases with height, showing
the pattern of the outward-running wave clearly seen in many roAp
stars. The high-frequency models predict a substantial difference
in the phase of REE lines formed at different layers in the upper
atmosphere. The low-frequency models often demonstrate a slow
height variation of the pulsation phase, followed by a more rapid
changes in the higher layers. This is reminiscent of the
standing-like followed by the running-like wave behaviour reported
in observational studies.

For a few roAp stars the radial velocity studies of different REEs
reveal a sudden drop of the pulsation amplitude and a $\pi$ radian
change in phase at high atmospheric layers. Occasionally, such
high-lying node surfaces in the vertical velocity appear in our
simulations at high magnetic latitudes. These nodes are produced
due to a superposition of the fast and slow magneto-acoustic modes
that attain similar amplitudes sufficiently far from the pulsation
poles. An example of the vertical velocity variation at the
magnetic latitude corresponding to $\psi=62.7$\degree\ in the
simulations with $B_{\rm d}=3$~kG showing a high-lying node is
given in Fig.~\ref{fig:example}. Though such kind of nodes are
quite persistent at high latitudes, they are not retained in the
integrated velocity data. It happens because the amplitude of
pulsations is small far from the poles for the assumed mode
geometry. However, we can speculated that the upper atmospheric
nodes can become visible in the radial velocity data due to an
inhomogeneous distribution of chemical elements, located in spots
close to the magnetic equator. Additionally, contribution of these
regions to the disk-integrated signal could be amplified by the
magnetic distortion of the mode geometry \citep{Saio2005}.

The non-detection of the pulsational variations of magnetic field
is in line with the results of our numerical simulations. The
disk-integrated amplitudes of the fluctuations of the field
modulus and longitudinal field are well below the current
detection limits attained in the time-resolved spectroscopic and
spectropolarimetric observations of roAp stars. Furthermore, our
models show that the depth-dependence of the magnetic field
variations is opposite to that of the radial velocity. Minute
fluctuations of magnetic field rapidly decrease above
$\log\tau_5=0$. Thus, an eventual detection of the pulsational
variation of magnetic field is far more likely using the lines
formed in deep layers, characterized by very weak or undetectable
velocity oscillations.

Finally, theoretical pulsation models computed in our study allow
us to comment on the REE line profile variation (LPV) of the REE
lines forming in the upper atmospheres of roAp stars. This
interesting topic was explored in several recent theoretical and
observational studies. With the aim to explain the anomalous
asymmetric LPV of strong REE lines, \citet{shibahashi:2008}
proposed a shock-wave model according to which the intrinsic
pulsation velocity amplitude at the REE line formation heights
($\log\tau_5=-4:-5$) reaches values comparable to the sound speed.
To retain an agreement with the observed substantially subsonic,
sinusoidal radial velocity curves, Shibahashi et al. postulated
that the REE line formation zone spans over a shock-wave train, so
that the observed signal corresponds to the vertically-averaged
pulsational variation. Our hydrodynamical models do not confirm
this picture. Apart from a moderate non-linearity in the uppermost
layers, the velocity variations are harmonic and their amplitudes
are well below the sound speed. For a broad range of the magnetic
inclinations ($\psi=0-80$\degree)  the REE line-forming zone above
$\log\tau_5=-2$ accommodates no more than one vertical pulsation
wavelength. For very high field inclinations ($\psi>80$\degree)
multiple nodes can appear in the vertical velocity but the
contribution of the corresponding surface regions to the
disk-integrated signal is negligible.

An alternative phenomenological model of the pulsational LPV of
REE lines \citep{kochukhov:2007} provides a closer match to the
behaviour of our theoretical simulations. Kochukhov et al. showed
that the observed LPV can be reproduced by a combination of the
subsonic velocity changes and the line width variation. The latter
can be related to the temperature fluctuation in the uppermost
layers (Fig.~\ref{fig:curves}).
However, detailed spectrum synthesis is required to study in
detail the LPV predicted by our models and to compare these
predictions with the observed pulsational line behaviour.

Detailed spectroscopic analysis and the comparison of the spectra
of different chemical elements synthesized in our simulations with
observations will be the aim of our future investigations.

\acknowledgements  EK is grateful to M. Collados for the helpful
comments of the manuscript. This research has been funded by the
Spanish Ministerio de Educaci{\'o}n y Ciencia through projects
AYA2007-63881 and AYA2007-66502. OK is a Royal Swedish Academy of
Sciences Research Fellow supported by a grant from the Knut and
Alice Wallenberg Foundation.


\end{document}